# Infection dynamics for fluctuating infection or removal rates regarding the number of infected and susceptible individuals

Seong Jun Park[a,b,*] , Seong-Geun Jeong[c] and M.Y. Choi[b]

[a] Global Science Research Center for Systems Chemistry, Chung-Ang University, Seoul 06974, Republic of Korea

[b] Department of Physics and Astronomy and Center for Theoretical Physics, Seoul National University, Seoul 08826, Republic of Korea

[c] Institute of Chemical Processes, Seoul National University, Seoul, 08826, Republic of Korea

## Abstract

In general, the rates of infection and removal (whether through recovery or death) are nonlinear functions of the number of infected and susceptible individuals. One of the simplest models for the spread of infectious diseases is the SIR model, which categorizes individuals as susceptible, infectious, recovered or deceased. In this model, the infection rate, governing the transition from susceptible to infected individuals, is given by a linear function of both susceptible and infected populations. Similarly, the removal rate, representing the transition from infected to removed individuals, is a linear function of the number of infected individuals. While nonlinear infection and removal rates have been extensively studied in deterministic epidemiological models, analytic results for stochastic dynamics with general nonlinear rates remain limited. This work presents an analytic expression for the number of infected individuals considering nonlinear infection and removal rates. In particular, we examine how the number of infected individuals varies as cases emerge and obtain the expression accounting for the number of infected individuals at each moment. This work paves the way for new quantitative approaches to understanding infection dynamics.

Email addresses: hagnos@snu.ac.kr (Seong-Geun Jeong); tcpsj123@cau.ac.kr (Seong Jun Park);
mychoi@snu.ac.kr (M.Y. Choi)

## 1. Introduction

Since the beginning of human history, infectious diseases caused by pathogens have posed a persistent threat. Viruses, which exist at the boundary between living and non-living entities, are microscopic infectious agents that replicate only within host organisms. Approximately 270 virus species are estimated to infect humans, causing over 200 diseases (Evans, 1976; Flint et al., 2020; Forni et al., 2022; L. Liu, 2014). Furthermore, illnesses caused by viruses are responsible for millions of mortalities on an annual basis (Carroll et al., 2018; Evans, 1976). The proliferation of new virus species, driven by technological advancements, has led to a steady rise in the number of identified viruses (Harko et al., 2014; Hethcote, 2000; Kermack & Mckendrick, 1927). The susceptibility of humans to viral infections is multifaceted, influenced by factors such as high population density, deforestation, and transportation infrastructure.

The analysis of interhost infection dynamics aims to determine the number of infected individuals. This analysis is instrumental in predicting herd immunity and formulating effective quarantine measures, such as social distancing. A substantial body of research has been dedicated to the development and application of mathematical models in epidemiology. Mathematical modeling of infectious disease spread involves categorization of the population, with individuals capable of moving between these categories.

One of the most straightforward compartment models is the susceptible, infectious, and recovered (SIR) model, which includes the numbers of susceptible, infectious, and recovered or deceased individuals (Harko et al., 2014; Hethcote, 2000; Kermack & Mckendrick, 1927; Kröger & Schlickeiser, 2020). In the SIR model, the total population, the sum of the three compartments, is fixed. Numerous studies have been conducted on modifications and extensions of the SIR model, such as models where there is no immunity after recovery (SIS model) (Hethcote, 1989; Kuhl, 2021), where immunity lasts only for a short period (SIRS) (Brauer & Castillo-Chavez, 2013; Cai et al., 2015; Kermack & McKendrick, 1933), where there is a latent period of the disease when the person is not infectious (SEIS and SEIR; E: exposed) (Anderson & May, 1991; Bianchi et al., 2019; Brauer & Castillo-Chavez, 2013; Kermack & Mckendrick, 1927), where infants may be born with immunity (MSIR; M: maternal immunity) (Brauer & Castillo-Chavez, 2013; Hethcote, 2000), and where the immunity wanes over time after infection (SEIRS) (Bjørnstad et al., 2020; Brauer & Castillo-Chavez, 2013; Hethcote, 2000). Many compartment models have been developed in epidemiology to estimate the total number of infected and recovered individuals (Allen et al., 2008; Gao et al., 2007; Kastalskiy et al., 2021; Keeling & Rohani, 2008; Paxson & Shen, 2022).

Viral infection and removal (recovery or death) rates are linear functions of the number of compartments (S or I) in the aforementioned models of disease spread. However, infection, recovery, and mortality rates may not be linear functions of the number of susceptible or removed individuals because each time a virus infection occurs, the rates vary with many factors, such as individual character, social environment, infection route (e.g., contact or droplet), and intrinsic properties of the virus. Several studies have focused on viral infection dynamics involving specific nonlinear infection and removal rates(Brauer & Castillo-Chavez, 2013; Calvo-Monge et al., 2023; Hethcote, 2000; Rohith & Devika, 2020). Nonlinear infection (incidence) rates have been widely studied in epidemiological models since the early work of (Capasso & Serio, 1978) and subsequent developments (Hethcote, 2000; W. Liu et al., 1987; Xiao & Ruan, 2007). These studies introduced various functional forms, including saturated and behavioral incidence rates, to capture realistic contact processes. However, most of these studies are formulated within deterministic compartment models and focus on specific functional forms of the infection and removal rates. Stochastic epidemic models have also been extensively investigated (Allen, 2008; Bailey, 1975; Nåsell, 2002), primarily focusing on Markovian dynamics with linear or specific rate structures. Most analytic results in these frameworks rely on linear or otherwise restricted rate functions, which limit their applicability to systems with more complex, interaction-driven dynamics. Analytic results for stochastic infection dynamics with arbitrary nonlinear infection and removal rates remain limited.

In this work, we address this gap by developing a general analytic framework for stochastic infection dynamics with arbitrary nonlinear infection and removal rates. Our approach is based on a birth-death process, where infection and removal events are respectively modeled as birth and death. We derive explicit expressions for the time-dependent probability distribution of the number of infected individuals, as well as for key summary quantities such as the mean population size.

Park and Choi (Park & Choi, 2023) studied a birth-death process in which birth and death rates depend on the population size of a product or customer undergoing the birth-death process.

Their proposed framework derives product number counting statistics to extend the framework presented in (Park & Choi, 2023) to the disease spread model.

This paper discusses the number of compartments (susceptible, infected, recovered or deceased individuals) when the infection and removal (recovery or death) rates are nonlinear functions of the number of infected and susceptible individuals. This work derives analytic expressions for the numbers of susceptible, infected, and recovered or deceased individuals when the infection and removal rates are nonlinear functions of the numbers of infected and removed (recovered or deceased) cases and that the sum of the three compartments (susceptible, infected, and removed individuals) is fixed. In real-world disease spread, the number of infected individuals increases due to susceptible individuals becoming infected and the arrival of externally infected individuals. Therefore, we provide an analytic expression for the number of infected and removed cases, accounting for the arrival of infected individuals.

The computational simulations prove that the analytical results are correct. We use the term "nonlinear infection or removal rates" for nonlinear functions of the number of susceptible and infected individuals. The main contributions of this study are analytical expressions for the number of individuals in three compartments (susceptible, infected, and removed) based on infection and removal rates as arbitrary functions of the number of susceptible or infected individuals.

This paper is organized as follows: Section 2 explains nonlinear infection and removal rates, providing analytical expressions for the number of individuals in the susceptible, infected, and removed compartments. This work assumes that the total population remains constant, that removed individuals (recovered or deceased) cannot be reinfected, and that there is no influx of infected individuals. Next, Section 3 describes the effect of the arrival of infected individuals on the numbers of susceptible, infected, and removed individuals, presenting the analytical expression of the populations of the three compartments. Finally, Section 4 summarizes and discusses the work.

## 2. Infection dynamics without the arrival of infected individuals

We suggest a new approach to analyzing and predicting the number of susceptible, infected, and removed individuals under nonlinear infection and removal rates. From an infected individual's viewpoint, infectious disease spread is a birth-death process because the infection corresponds to the birth of an infected case, whereas removal corresponds to its death. The infection rate is a function of the number of susceptible and infected individuals, and the removal rate is a function of the number of infected individuals. Hence, this work considers a birth-death process where the birth and death rates depend on the number of subjects undergoing the birth-death process and the number of subjects interacting with the subjects of interest.

This work obtains the analytic expression of the mean number of subjects undergoing a birth-death process whose birth and death rates are arbitrary functions of the number of subjects involved. The mean number of subjects undergoing the birth-death process is the number of infected individuals because the mean *E*(*X*) is the value of *b* that minimizes $E\left((X-b)^2\right)$, where *E* is the mathematical expectation, *X* denotes a random variable, and *b* is not a function of *X* (Hogg et al., 2013).

Next, this work examines the birth-death process described by $A \longrightarrow B \longrightarrow X$, where *B* undergoes birth and death. The birth rate (the transition rate from *A* to *B*) depends on *A* and *B*; the death rate (the transition rate from *B* to *X*) is a function of *B* alone. This work assumes that the total population (the sum of populations *A*, *B*, and *X*) is a constant, *C*. Initially, there are $C - 1$ members in population *A*, one member in population *B*, and no member in population *X*. This birth-death process is similar to the spread of an infectious disease: Compartments *A*, *B*, and *X* correspond to the sets of susceptible, infected, and removed individuals, respectively. In general, the infection rate depends on the number of susceptible and infected individuals, and the removal rate depends on the number of infected individuals. When studying the time evolution of birth-death processes, the most common objective is determining the probability $P_n(t)$ of *n* members in compartment *B* at time *t*. Since *B* corresponds to infected individuals, $P_n(t)$ is the probability that there are *n* infected individuals at time *t*. When the numbers of susceptible and infected individuals are *s* and *n*, respectively, the time until the following infection occurs is an exponential random variable with rate $\lambda_{n,s} (\geq 0)$. During an infection, the number of infected cases increases by one (changing from *n* to $n + 1$), and the number of susceptible individuals decreases by one (from $s$ to $s - 1$). When the infection count exceeds one ( $n \geq 1$), the time until the subsequent removal is an exponential random variable with rate $(\mu_n \geq 0)$. At death or recovery, the number of infected cases decreases by one while the number of removed individuals increases by one. In realistic interhost infections, the infection and removal rates may depend not only on the numbers of infected and susceptible individuals but also on time. We rewrite the rates as $\lambda_{n,s}(t)$ and $\mu_n(t)$. In this work, we focus on infection and removal rates that depend only on the number of infected and susceptible individuals. Extensions to time- and compartment-size- dependent rates are left for future work.

We now obtain the probability $P_n(t)$ as a function of the infection rate $\lambda_{n,s}$ and the removal rate $\mu_n$, based on a disease spread model that satisfies the following three assumptions. First, the total population (the sum of the susceptible, infected, and removed populations) is constant $C$. Second, removed individuals cannot be reinfected. Third, at $t = 0$, the number of susceptible individuals is $C - 1$, with one index case and no individuals removed (recovered or deceased). To our knowledge, no analytic expression of $P_n(t)$ exists for infection and removal rates given by nonlinear functions of the numbers in compartments in this infectious disease spread model. To derive $P_n(t)$, we introduce the probability density function of time to complete $m$ infections with $n$ infected individuals at the instant of the $m$th infection, denoted by $f_{n,m}(t)$. Under these assumptions, $P_n(t)$ is written as follows (see the Supplementary Material):

$$\begin{cases} P_0(t) = \int_0^t \left( \mu_1 e^{-(\mu_1 + \lambda_{1,S_0})\tau} + \sum_{m=1}^{S_0} \sum_{k=2}^{m+1} \int_0^\tau f_{k,m}(\tau_1) v_{k,m,S_0}(\tau - \tau_1) d\tau_1 \right) d\tau \\ P_n(t) = \sum_{m=n}^{S_0} \frac{f_{n+1,m}(t)}{\lambda_{n,S_0-(m-1)}} \quad n = 1, 2, \cdots, C-1 \\ P_C(t) = \int_0^t f_{C,S_0}(\tau) e^{-\mu_C(t-\tau)} d\tau \end{cases}, \qquad (1)$$

where $f_{k,m}(t)$ $\left(m = 1, 2, \cdots, S_0; k = 2, 3, \cdots, m+1\right)$ satisfies

$$
\begin{aligned}
f_{k,m}(t) &= \int_0^t f_{k-1,m-1}(\tau)\lambda_{k-1,S_0-(m-1)} e^{-(\lambda_{k-1,S_0-(m-1)}+\mu_{k-1})(t-\tau)} d\tau \\
&+ \sum_{j=0}^{m-k} \left\{ \left( \prod_{l=0}^{j} \mu_{k+j-l} \right) \int_0^t f_{k+j,m-1}(\tau_1) \int_0^{t-\tau_1} g_{k,j,S_0,m}(\tau_2) \lambda_{k-1,S_0-(m-1)} e^{-(\lambda_{k-1,S_0-(m-1)}+\mu_{k-1})(t-\tau_1-\tau_2)} d\tau_2 d\tau_1 \right\}
\end{aligned} \tag{2}
$$

with $g_{k,j,S_0,m}(t)$ representing the inverse Laplace transform of

$\hat{g}_{k,j,S_0,m}(s) = \int_0^\infty g_{k,j,S_0,m}(t) e^{-st} dt = \prod_{l=0}^{j} \left( s + \mu_{k+j-l} + \lambda_{k+j-l,S_0-(m-1)} \right)^{-1}$, and $v_{k,m,S_0}(t)$ denotes the inverse Laplace transform of $\hat{v}_{k,m,S_0}(s) = \int_0^\infty v_{k,m,S_0}(t) e^{-st} dt = \prod_{l=1}^{k} \frac{\mu_l}{s + \mu_l + \lambda_{l,S_0-m}}$. The function $f_{k,m}(t)$ can be interpreted as the probability density that the *m*-th infection occurs at time *t*, with *k* infected individuals present at that moment. Thus, it encodes the temporal sequence of infection events. The integral structure of $f_{k,m}(t)$ reflects a summation over all possible sequences of infection and removal events: to have *k* infected individuals at time *t* when the m-th infection occurs, the first term in Eq. (2) represents the case where there are *k*-1 infected individuals at the time $\tau$ of the (*m*-1)-th infection, and one additional infection occurs without any removal during the interval $t-\tau$. The second term accounts for cases where there are *k+j* infected individuals at the (*m*-1)-th infection at time $\tau_1$, followed by *j*+1 removal events during the interval $\tau_2$, and then one infection occurs without removal during $t-\tau_1-\tau_2$. Equation (2) presents a hierarchical structure, where $f_{k,m}(t)$ is determined by $f_{1,m-1}(t), f_{2,m-1}(t), \cdots,$ and $f_{k-1,m-1}(t)$. Thus, we can evaluate $f_{k,m}(t)$ only if $f_{2,1}(t)$ is known. Under the initial condition (the third assumption), we have $f_{2,1}(t) = \lambda_{1,S_0} e^{-(\lambda_{1,S_0}+\mu_1)t}$ because the first infection occurs at a rate of $\lambda_{1,S_0}$. Moreover, the index case recovery or death must not occur before the first infection. We can see from Eq. (2) that $f_{k,m}(t)$ is always positive, since $\lambda_{n,s}$ and $\mu_n$ are positive. From Eq. (1), the positivity of $f_{k,m}(t)$, $\lambda_{n,s}$ and $\mu_n$ ensures that $P_n(t)$ is also positive. We now examine the physical interpretation of Eq. (1). First, we consider its first equation: the first integral represents the case where the index case is removed before any infection occurs, while the second integral accounts for the removal of all infected individuals after at least one infection has occurred. Next, we turn to the second equation, where the term $\dfrac{f_{n+1,m}(t)}{\lambda_{n,S_0-(m-1)}}$ represents the contribution to $P_n(t)$ for $n=1,2,\cdots,C-1$. Finally, we consider the last equation. It shows that, to have C infected individuals at time *t*, the system already C infected individuals at an earlier time, with the required number ($S_0$) of infection events, and no removal occurring during the intervening period.

Equation (1), which describes how the number of infected individuals varies in time, manifests that $P_n(t)$ satisfies the normalization condition: $\sum_{n=0}^{C} P_n(t)=1$. Differentiating Eq. (1), we obtain the foundational equation, i.e., the differential-difference equation governing the time evolution of a birth-death process in this infectious disease spread model:

$$
\begin{cases}
\dfrac{\partial P_0(t)}{\partial t} = \mu_1 P_1(t) \\
\dfrac{\partial P_1(t)}{\partial t} = -\sum_{m=1}^{S_0} f_{2,m}(t) - \mu_1 P_1(t) + \mu_2 P_2(t) \\
\dfrac{\partial P_n(t)}{\partial t} = -\sum_{m=n}^{S_0} f_{n+1,m}(t) - \mu_n P_n(t) + \sum_{m=n-1}^{S_0} f_{n,m}(t) + \mu_{n+1} P_{n+1}(t) \qquad n=2,3,\cdots,C-1 \\
\dfrac{\partial P_C(t)}{\partial t} = f_{C,S_0}(t) - \mu_C \int_0^t f_{C,S_0}(\tau) e^{-\mu_C(t-\tau)} d\tau
\end{cases}
. \tag{3}
$$

The structure of Eq. (3) resembles differential–difference equations arising in stochastic epidemic models and branching processes (Allen, 2008; Harris, 1963). However, classical branching processes typically assume independent individuals and linear transition rates. In contrast, the present formulation incorporates nonlinear infection and removal rates that depend on both susceptible and infected populations, as well as a finite population constraint. Therefore, our approach can be viewed as a generalization of branching process models to interacting systems with arbitrary rate functions.

Next, we derive the numbers of susceptible, infected, and removed individuals using Eq. (1), where the infected population is a random variable. The mean number of infections obtains the form

$$\langle n(t)\rangle = \sum_{n=0}^{C} nP_n(t) = \sum_{n=1}^{S_0}\sum_{m=n}^{S_0} \frac{nf_{n+1,m}(t)}{\lambda_{m,S_0-(m-1)}} + C\int_0^t f_{C,S_0}(\tau)e^{-\mu_C(t-\tau)}d\tau\ , \tag{4}$$

where $f_{n,m}(t)$ is given by Eq. (2). The first term of the right-hand side of eq. (4) is derived from $\sum_{n=0}^{C-1} nP_n(t)$ and the last term is equivalent to $CP_C(t)$. Given a birth-death process, removed (recovered or deceased) individuals undergo only one birth. With the removal (recovery or death) rate $\mu_n$, the recovery count and death toll $r(t)$ thus reads (Park & Choi, 2022; Ross, 2019):

$$\langle r(t)\rangle = \int_0^t \langle \mu_n(\tau)\rangle d\tau\ , \tag{5}$$

where $\langle \mu_n(t)\rangle = \sum_{n=0}^{C} \mu_n P_n(t)$ is the mean removal rate. In terms of removed individuals, this infection model is identical to pure birth process. Thus, the number of removed individuals is the integral of the mean removal rate with respect time. Then the number $s(t)$ of susceptible individuals in the total population $C$ is given by

$$\langle s(t)\rangle = C - \langle n(t)\rangle - \langle r(t)\rangle\ . \tag{6}$$

In Eqs. (4) to (6), the number in each compartment must be computed even if the infection and removal rates are arbitrary functions of infected and susceptible populations. The positivity of $P_n(t)$ makes it clear that the numbers of infected, susceptible, and removed individuals remain positive for all time $t$. Figure 1 presents the number of susceptible, infected, and removed individuals for several rates of infection and removal, demonstrating that Eqs. (4) to (6) agree with the simulation results.

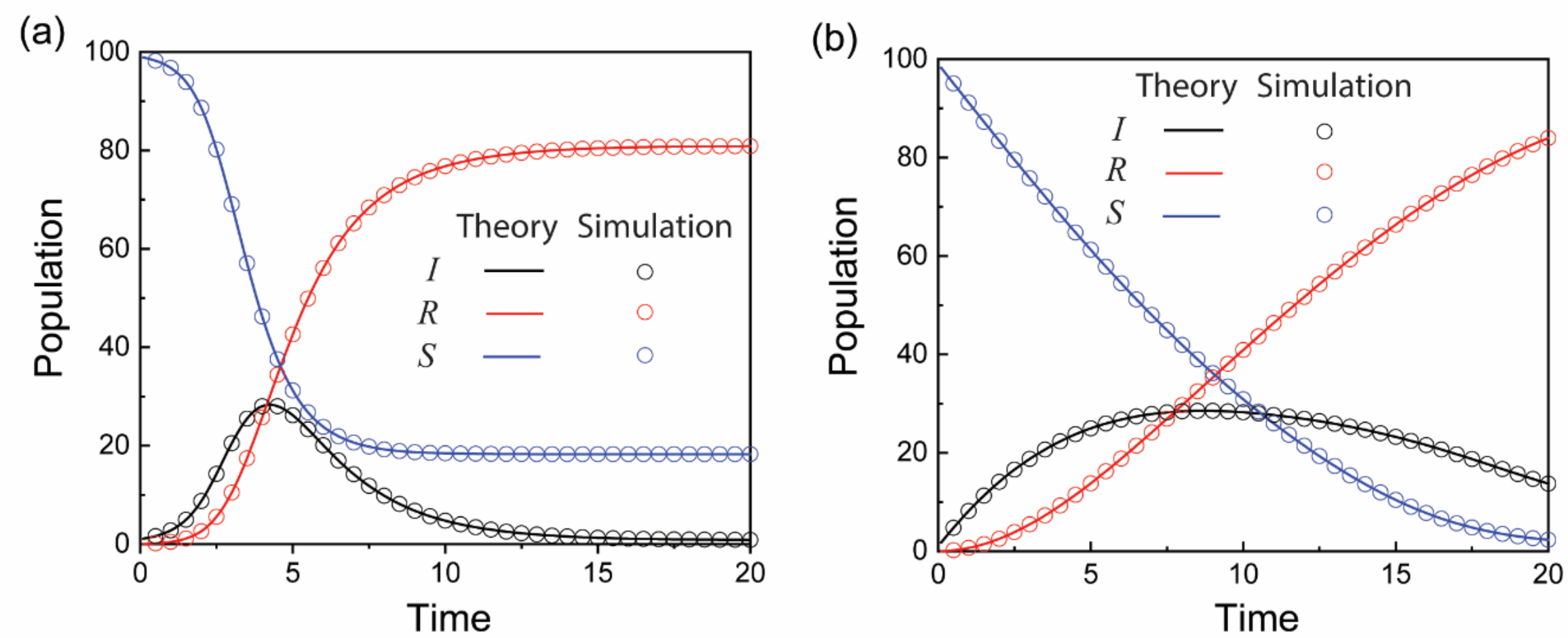

Figure 1. Infected (I), removed (R), and susceptible (S) populations with nonlinear infection and removal rates without the arrival of infected cases. Circles and lines plot data from simulations and theoretical results from Eqs. (4) to (6), respectively, for rates (a) $\lambda_{n,s} = 0.07n^{1.2}s^{0.6}$ and $\mu_n = 0.2n^{1.3}$; (b) $\lambda_{n,s} = 0.2(n+s)^{0.8}$ and $\mu_n = 0.1n^{1.2}$.

The compartment sizes are obtained through a sequence of steps. The derivation of the analytic expressions for the susceptible, infected, and removed populations is summarized schematically in Fig. 2.

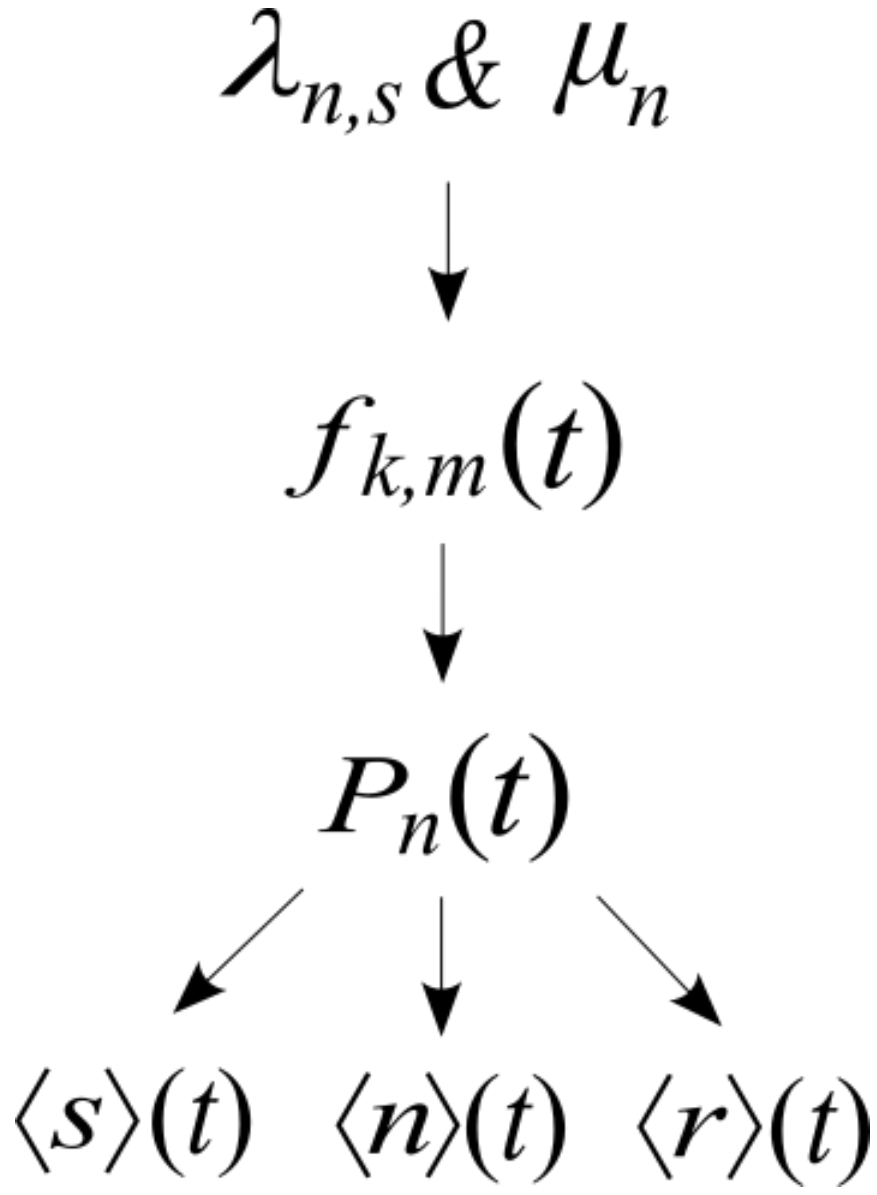

Figure 2. Schematic overview of the analytic framework for infection dynamics without external arrivals. This scheme illustrates the derivation from nonlinear infection and removal rates to the number of susceptible, infected, and removed individuals. The infection rate $\lambda_{n,s}$ increases the number of infected individuals from $n$ to $n$+1 while decreasing the number of susceptible individuals from $s$ to $s$-1. The removal rate $\mu_n$ decreases the number of infected individuals by one and increases the number of removed (recovered or deceased) individuals by one. The function $f_{k,m}(t)$ (Eq. (2)) is the probability density for the time at which m infections have occurred, with $k$ infected individuals present at that moment. Equation (1) gives $P_n(t)$, the probability of having $n$ infected individuals at time $t$. Equations (4)–(6) provide the numbers of infected, removed, and susceptible individuals, respectively.

It is of interest to compare Eq. (3) with the number of each compartment in the SIR model. The SIR model is described by the following system of ordinary differential equations:

$$\begin{cases} \dfrac{dS(t)}{dt} = -\beta I(t)S(t) \\ \dfrac{dI(t)}{dt} = \beta I(t)S(t) - \gamma I(t) \\ \dfrac{dR(t)}{dt} = \gamma I(t) \end{cases} \tag{7}$$

with positive constants $\beta$ and $\gamma$, where $S(t)$, $I(t)$, and $R(t)$ denote the number of susceptible, infected, and removed (recovered or deceased) individuals, respectively, at time $t$. Multiplying both sides of Eq. (3) by $n$ and summing over $n$ ($= 0, 1, 2, \cdots, C$) yield the rate of change in the number of infected individuals, which differs from $dI(t)/dt$ given in Eq. (7). Accordingly, Eqs. (4) to (6) do not agree with the solutions of the set of ordinary differential equations describing the SIR model. Figure 3 presents the discrepancy between the results from Eqs. (4) to (6) and the solutions of Eq. (7).

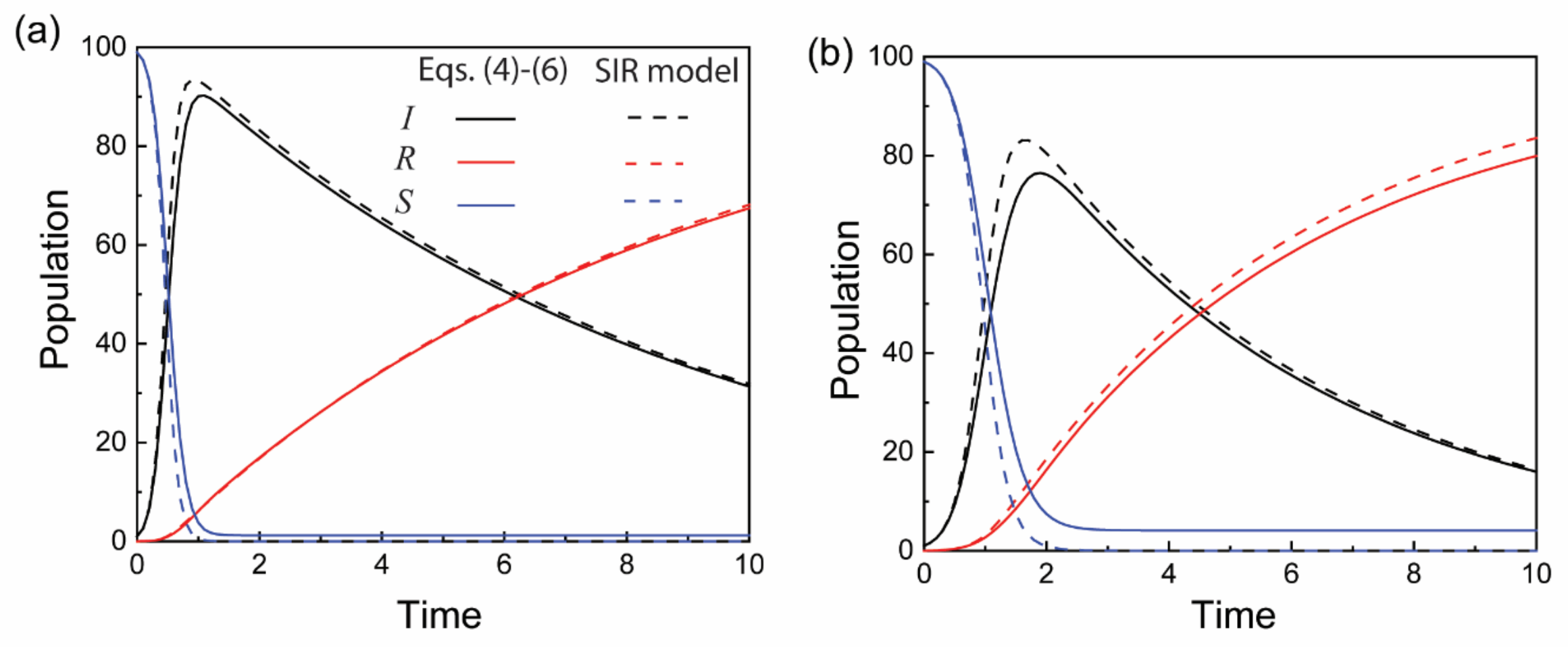

Figure 3. Infected (I), removed (R), and susceptible (S) populations resulting from Eqs. (4) to (6) (solid lines) and from the SIR model described by Eq. (7) (broken lines). Rates are given by (a) $\left(\lambda_{n,s},\mu_n\right)=\left(0.1ns,0.12n\right)$ in Eqs. (4) - (6) and $\left(\beta,\gamma\right)=\left(0.1,0.12\right)$ in Eq. (7);

(b) $\left(\lambda_{n,s},\mu_n\right)=\left(0.05ns,0.2n\right)$ in Eqs. (4) - (6) and $\left(\beta,\gamma\right)=\left(0.05,0.2\right)$ in Eq. (7).

To illustrate the practical impact of nonlinear infection rates, we include a representative comparison between a nonlinear rate and its corresponding linear approximation. Specifically, we consider a nonlinear infection rate of the form ( $\lambda_{n,s}=\dfrac{\beta ns}{1+\alpha n}$ ) and compare it with the standard linear rate ( $\lambda_{n,s}=\beta ns$ ) in Fig. 4, while modeling the removal rate as $\mu_n=\gamma n$ . The nonlinear infection rate ( $\lambda_{n,s}=\dfrac{\beta ns}{1+\alpha n}$ ) captures saturation effects due to limited contact opportunities or behavioral changes at high infection levels. The results show that the nonlinear rate leads to slower growth and a reduced peak in the number of infected individuals compared to the linear approximation. This difference arises from the effective suppression of infection at higher population levels, which cannot be captured by linear models. This example demonstrates that allowing nonlinear infection rates can significantly alter the temporal dynamics of the system.

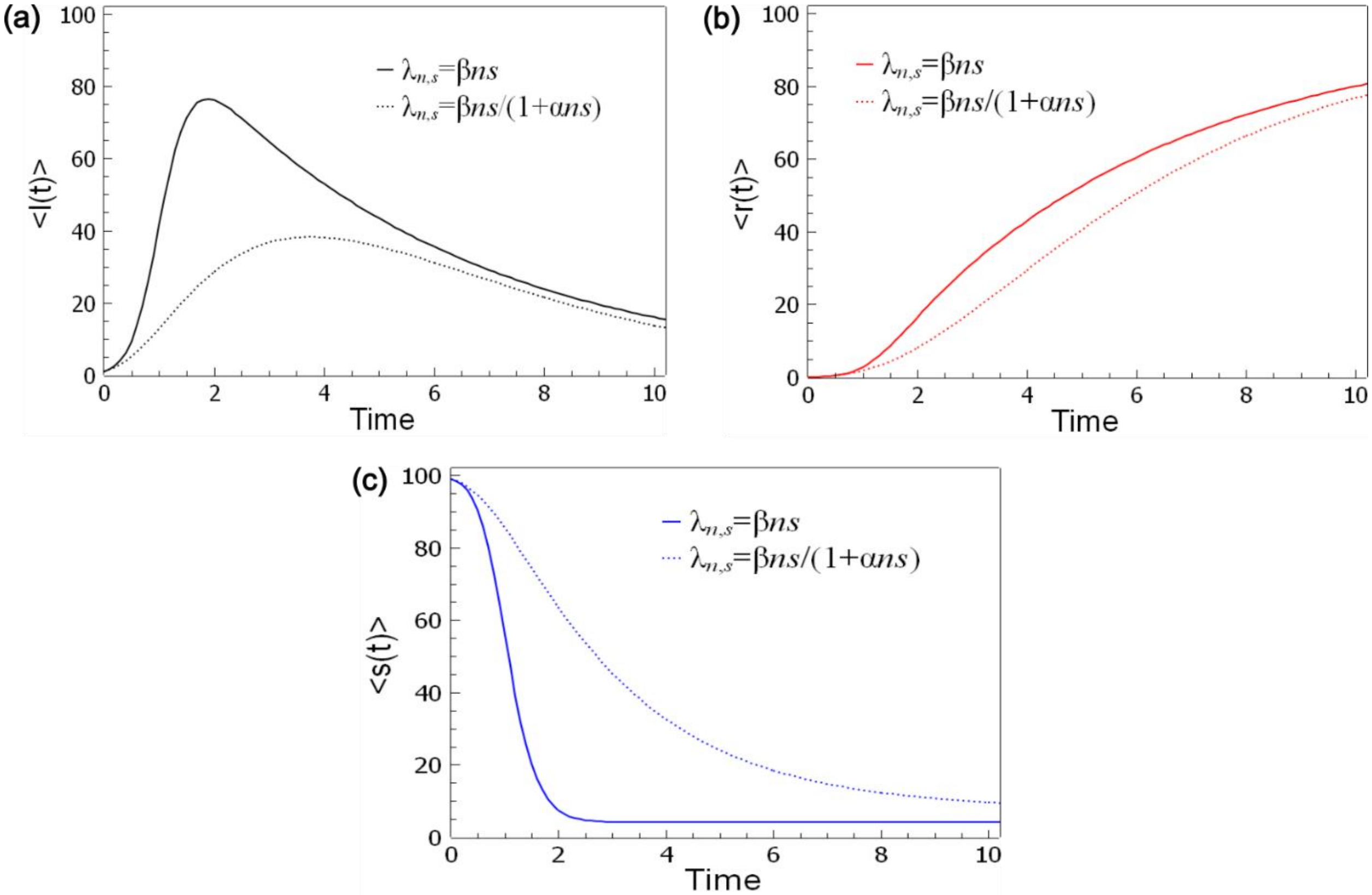

Figure 4. Comparison of infection dynamics under a nonlinear infection rate ( $\lambda_{n,s} = \dfrac{\beta ns}{1+\alpha n}$ ) and its linear approximation ( $\lambda_{n,s} = \beta ns$ ). The removal rate is $\mu_n = \gamma n$ . The parameter values are $\beta = 0.05$ , $\alpha = 1$ , and $\gamma = 0.2$ . The solid and dotted lines show compartment sizes under the linear and nonlinear infection rates, respectively, computed from Eqs. (4)–(6). (a) the number of infected individuals; (b) the number of removed individuals; (c) the number of susceptible individuals.

## 3. Infection dynamics with the arrival of infected individuals

In many real-world situations where an infectious disease is spreading, the number of infected individuals may increase due to the arrival of infected individuals. Therefore, the increase in the number of infected cases can be divided into two types: infections started by the index case and the influx of infected cases. This work addresses these two types of increases in infected cases. The infectious disease spread model studied is the same as in Section 2, except for the arrival of infected individuals.

This work focuses on the current number of infected individuals. Suppose the number of infections introduced is *v*, occurring at $a_1, a_2, \cdots, a_v$. Starting with one infected individual and $S_0$ susceptible individuals, the total population *C* reaches $S_0 + v + 1$ after time $a_v$ when the last introduced case occurs. As the infected population increases due to infections and arrivals, instead of $f_{n,m}(t)$, we introduce $f_{k,m,r}(t)$, the probability density of the time to complete *m* infections and *r* arrivals with *k* infected individuals at this time, where *k* runs from 1 to *m*+*r*+1 ($k = 1, \cdots, m + r + 1$), *m* from 0 to $S_0$ ($m = 0, 1, \cdots, S_0$), and *r* from 0 to *v* ($r = 0, 1, \cdots, v$). Under the initial condition $P_n(0) = \delta_{n1}$, we write $f_{k,m,r}(t)$ as follows:

$$f_{k,m,r}(t) = \begin{pmatrix} \int_0^t f_{k-1,m-1,r}(\tau_1) \lambda_{k-1,S_0-(m-1)} e^{-(\lambda_{k-1,S_0-(m-1)} + \mu_{k-1})(t-\tau_1)} d\tau_1 \\ + \sum_{q=0}^{m+r-k} \int_0^t f_{k+q,m-1,r} \int_0^{t-\tau_1} g_{k,q,S_0,m}(\tau_2) \lambda_{k-1,S_0-(m-1)} e^{-(\lambda_{k-1,S_0-(m-1)} + \mu_{k-1})(t-\tau_1-\tau_2)} d\tau_2 d\tau_1 \end{pmatrix} \theta(a_{r+1} - t)$$
$$+ \begin{pmatrix} \int_0^{a_r} f_{k-1,m,r-1}(\tau_1) e^{-(\lambda_{k-1,S_0-m} + \mu_{k-1})(a_r-\tau_1)} d\tau_1 \\ + \sum_{q=0}^{m+r-k} \int_0^{a_r} f_{k+q,m,r-1}(\tau_1) \int_0^{a_r-\tau_1} g_{k,q,S_0,m+1}(\tau_2) e^{-(\lambda_{k-1,S_0-m} + \mu_{k-1})(a_r-\tau_1-\tau_2)} d\tau_2 d\tau_1 \end{pmatrix} \delta(t - a_r) \quad , (8)$$

where $g_{k,q,S_0,m}(t)$ denotes the inverse Laplace transform of

$\hat{g}_{k,q,S_0,m}(s) = \int_0^\infty g_{k,q,S_0,m}(t) e^{-st} dt = \prod_{l=0}^{q} \left( s + \mu_{k+q-l} + \lambda_{k+q-l,S_0-(m-1)} \right)^{-1}$, $\delta(x)$ represents the Dirac delta function, and $\theta(x)$ denotes the Heaviside step function. The function $f_{k,m,r}(t)$ can be interpreted as the probability density that the *m*-th infection and *r* arrivals occur at time *t*, with *k* infected individuals present at that moment. It thus describes the temporal sequence of infection events. Its integral structure of $f_{k,m,r}(t)$ represents a sum over all possible sequences of infection, arrival and removal events. To have *k* infected individuals at time *t* when the m-th infection and the r-th arrival occur, two cases are possible. (i) there are *k*-1 infected individuals just before time *t*, following the (*m*-1)-th infection and the *r*-th arrival occur, and one infection occurs during the remaining time; or (ii) there are *k*-1 infected individuals after the m-th infection and the (r-1)-th arrival, and one arrival occurs during the remaining time. The first and second cases correspond to the first and second terms in Eq. (8), respectively. Equation (8) can only be evaluated if $f_{1,0,1}(t)$, $f_{2,0,1}(t)$, and $f_{2,1,0}(t)$ are known. In the case of $m = 0$ and $r = 1$, the index case is removed to have one infection count at time *t*, and the infection must not occur until time $a_1$. This case is expressed as $f_{1,0,1}(t) = \int_0^{a_1} \mu_1 e^{-(\mu_1 + \lambda_{1,S_0})\tau} d\tau \delta(t - a_1)$. If the index case survives and becomes infected during the time $a_1$, the number of infected individuals is two at time $a_1$. Then, we have $f_{2,0,1}(t) = e^{-(\mu_1 + \lambda_{1,S_0}) a_1} \delta(t - a_1)$. The case of $m = 1$ and $r = 0$ represents one infection, with no infected individuals arriving. There are two cases immediately after the first infection: namely, $f_{2,1,0}(t) = \lambda_{1,S_0} e^{-(\lambda_{1,S_0} + \mu_1)t} \theta(a_1 - t)$. For a detailed derivation of Eq. (8), see the Supplementary Material. We can see from Eq. (8) that $f_{k,m,r}(t)$ is always positive, since $\lambda_{n,s}$ and $\mu_n$ are positive.

The next step is to obtain $P_n(t)$. Equation (8), which accounts for the infection count at the time of infection and arrival, leads Eq. (9) to obtain the form

$$
\begin{cases}
P_0(t) = \int_0^t \left( \mu_1 e^{-(\mu_1 + \lambda_{1,S_0})\tau} \theta(a_1 - t) + \sum_{m=0}^{S_0} \sum_{\substack{r=0 \\ m+r \geq 1}}^{v} \sum_{k=1}^{m+r+1} \left( \int_0^{\tau} f_{k,m,r}(\tau_1) v_{k,m,S_0}(\tau - \tau_1) d\tau_1 \, \theta(a_{r+1} - t) \right) \right) d\tau \\
P_n(t) = \frac{1}{2} \sum_{m=0}^{S_0} \sum_{\substack{r=0 \\ m+r \geq n}}^{v} \left( \frac{x_{n+1,m,r}(t)}{\lambda_{n,S_0-(m-1)}} + y_{n+1,m,r}(t) \frac{\theta(a_r - t)}{\delta(t - a_r)} \right) \qquad n = 1, 2, \cdots, S_0 + v \\
P_{S_0+v+1}(t) = \int_0^t f_{S_0+v+1, S_0+1, v}(\tau) e^{-\mu_{S_0+v+1}(t-\tau)} d\tau
\end{cases}
\tag{9}
$$

with

$$
x_{k,m,r}(t) = \left( \begin{aligned} & \int_0^t f_{k-1,m-1,r}(\tau_1) \lambda_{k-1,S_0-(m-1)} e^{-(\lambda_{k-1,S_0-(m-1)} + \mu_{k-1})(t-\tau_1)} d\tau_1 \\ & + \sum_{q=0}^{m+r-k} \int_0^t f_{k+q,m-1,r} \int_0^{t-\tau_1} g_{k,q,S_0,m}(\tau_2) \lambda_{k-1,S_0-(m-1)} e^{-(\lambda_{k-1,S_0-(m-1)} + \mu_{k-1})(t-\tau_1-\tau_2)} d\tau_2 d\tau_1 \end{aligned} \right) \theta(a_{r+1} - t),
$$

$$
y_{k,m,r}(t) = \left( \begin{aligned} & \int_0^{a_r} f_{k-1,m,r-1}(\tau_1) e^{-(\lambda_{k-1,S_0-m} + \mu_{k-1})(a_r-\tau_1)} d\tau_1 \\ & + \sum_{q=0}^{m+r-k} \int_0^{a_r} f_{k+q,m,r-1}(\tau_1) \int_0^{a_r-\tau_1} g_{k,q,S_0,m+1}(\tau_2) e^{-(\lambda_{k-1,S_0-m} + \mu_{k-1})(a_r-\tau_1-\tau_2)} d\tau_2 d\tau_1 \end{aligned} \right) \delta(t - a_r),
$$

and $v_{k,m,S_0}(t)$ the same as that given below Eq. (2). From Eq. (9), the positivity of $f_{k,m,r}(t)$, $\lambda_{n,s}$ and $\mu_n$ ensures that $P_n(t)\,(n=0,1,2,\cdots,S_0+v+1)$ is also positive. We now consider its physical interpretation. In the first equation, the first integral represents the case where the index case is removed before any infection or arrival occurs, while the second integral accounts for the removal of all infected individuals after at least one infection or arrival has taken place. Next, in the second equation, the terms $\dfrac{x_{n+1,m,r}(t)}{\lambda_{n,S_0-(m-1)}}$ and $y_{n+1,m,r}(t)\dfrac{\theta(a_r-t)}{\delta(t-a_r)}$ represent contributions in which either an infection or an arrival is the final event before the number of infected individuals reaches *n*. Finally, the last equation shows that, to have $S_0+v+1$ infected individuals at time *t*, the system must already have $S_0+v+1$ infected individuals at an earlier time, with the required number ($S_0+v$) of infection events having occurred and no removal taking place during the intervening period. The differential-difference equation satisfied by Eq. (9) has a similar structure to that of Eq. (3):

$$
\begin{cases}
\dfrac{\partial P_0(t)}{\partial t} = \mu_1 P_1(t) \\
\dfrac{\partial P_1(t)}{\partial t} = -\displaystyle\sum_{m=0}^{S_0}\sum_{\substack{r=0\\ m+r\ge 1}}^{v} f_{2,m,r}(t) - \mu_1 P_1(t) + \mu_2 P_2(t) \\
\dfrac{\partial P_n(t)}{\partial t} = -\displaystyle\sum_{m=0}^{S_0}\sum_{\substack{r=0\\ m+r\ge n}}^{v} f_{n+1,m,r}(t) - \mu_n P_n(t) + \sum_{m=0}^{S_0}\sum_{\substack{r=0\\ m+r\ge n-1}}^{v} f_{n,m,r}(t) + \mu_{n+1}P_{n+1}(t) \qquad n=2,3,\cdots,S_0+v \\
\dfrac{\partial P_{C_v}(t)}{\partial t} = f_{C_v,S_0,v}(t) - \mu_{C_v}\displaystyle\int_0^t f_{C_v,S_0,v}(\tau)e^{-\mu_{C_v}(t-\tau)}d\tau
\end{cases}
\tag{10}
$$

Equation (9) yields the number of infected individuals as follows:

$$
\langle n(t)\rangle = \frac{1}{2}\sum_{n=1}^{S_0+v}\sum_{m=0}^{S_0}\sum_{\substack{r=0\\ m+r\ge n}}^{v} n\left(\frac{x_{n+1,m,r}(t)}{\lambda_{n,S_0-(m-1)}} + y_{n+1,m,r}(t)\frac{\theta(a_r-t)}{\delta(t-a_r)}\right) + (S_0+v+1)\int_0^t f_{S_0+v+1,S_0+1,v}(\tau)e^{-\mu_{S_0+v+1}(t-\tau)}d\tau\,.
$$

(11)

When there is no external infection arrival ($v = 0$), Eq. (11) reduces to Eq. (4), and Eq. (9) reduces to Eq. (1). The number of removed individuals has the same structure as Eq. (5):

$\langle r(t)\rangle = \int_0^t \langle \mu_n(\tau)\rangle d\tau$ , except for $\langle \mu_n(t)\rangle = \sum_{n=0}^{C_v} \mu_n P_n(t)$ with $P_n(t)$ given in Eq. (9). The total population, denoted by $C_v$ , fluctuates over time in such a way that

$$C_v = \begin{cases} S_0 + 1 & t < a_1 \\ S_0 + 1 + z & a_z < t \le a_{z+1} \ (z = 1, 2, \cdots, v-1) \\ S_0 + 1 + v & a_v \le t \end{cases},$$

which then gives the number of susceptible individuals, $\langle s(t)\rangle = C_v - \langle n(t)\rangle - \langle r(t)\rangle$ . Thus, we compute the number of susceptible, infected, and removed individuals. Even though the infection rate depends on the infection count and the number of recoveries or deaths, the removal rate depends only on the infection count. These results, where external infection arrival is accounted for, are confirmed through simulations, as shown in Fig. 5.

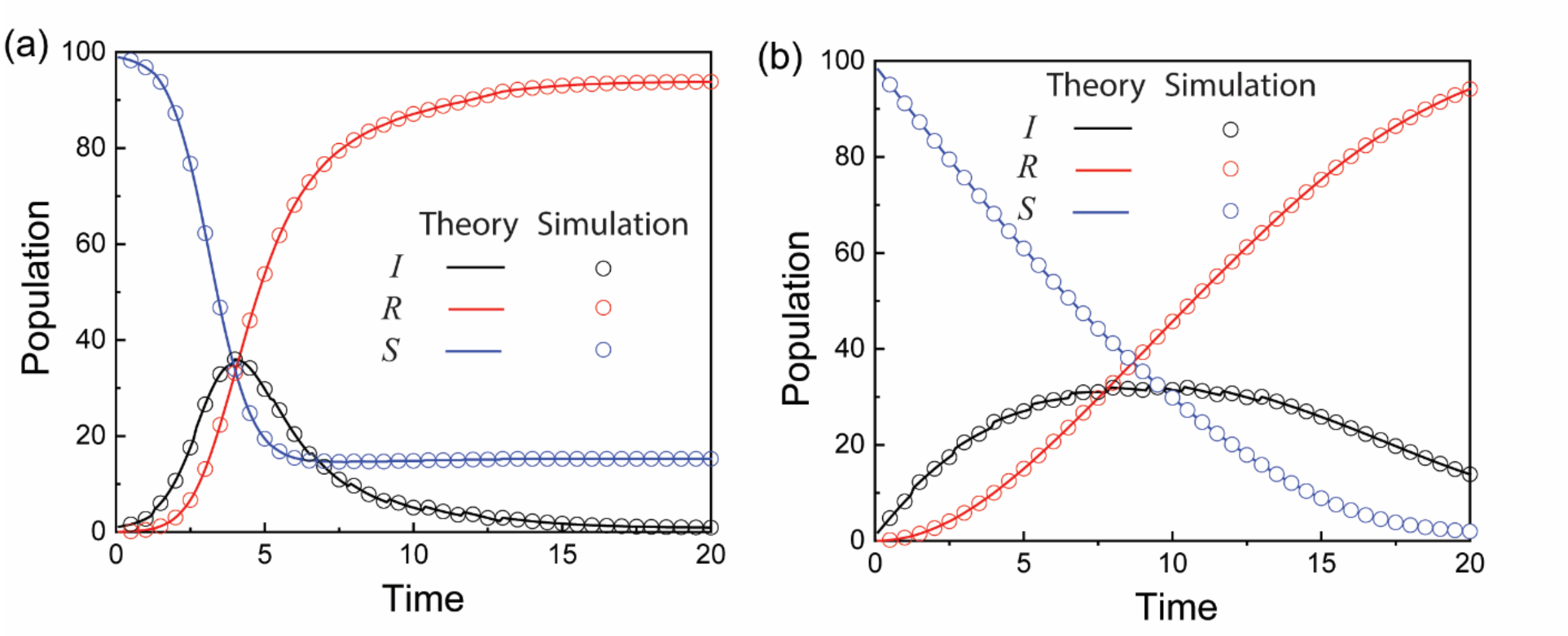

Figure 5. Infected (I), removed (R), and susceptible (S) populations with nonlinear infection and removal rates with the arrival of infected cases. The number of infected individuals introduced from outside is 10, and introduced infected cases occur every 1.3-time units. Circles and lines plot data from simulations and theoretical results from Eq. (11) and equations below, respectively, for rates (a) $\lambda_{n,s} = 0.07n^{1.2}s^{0.6}$ and $\mu_n = 0.2n^{1.3}$; (b) $\lambda_{n,s} = 0.2(n+s)^{0.8}$ and $\mu_n = 0.1n^{1.2}$ .

The compartment sizes with external arrivals are derived through a similar sequence of steps, with Eq. (8) used in place of Eq. (2).

## 4. Summary and discussion

This paper proposes a method of treating infectious disease spread viewed as a birth-death process, where birth corresponds to infection and death corresponds to removal (recovery or death). In this process, the birth/infection rate depends on the number of susceptible and infected individuals while the death/removal rate depends on the infected population. The mean number of subjects undergoing the birth-death process is considered to be the number of infected individuals because the mean represents the center of a collection of numbers. When $X$ is a random variable and $E(u(x))$ is the mathematical expectation of the function $u(x)$, then the value of $b$ for which $E((X-b)^2)$ is minimum is $E(X)$.

This work considers a simple model of infectious disease spread. The population is divided into three compartments: S (susceptible), I (infected), and R (removed). Individuals can transition between these states without becoming reinfected. The compartmental model is schematized as follows: $S \xrightarrow{\lambda_{n,s}} I \xrightarrow{\mu_n} R$, where $\lambda_{n,s}$ and $\mu_n$ denote the infection rate and the removal rate, respectively, depending on the number of susceptible or infected individuals. This work introduces the probability density function $f_{n,m}(t)$ of $m$ infections with $n$ infected individuals at time $t$, as given by Eq. (2). Based on $f_{n,m}(t)$, we have obtained the probability $P_n(t)$ in Eq. (1). The differential-difference equation governing the dynamics of the spread of an infectious disease is described in Eq. (3). Using Eq. (1), we have derived the analytic expressions of the numbers of susceptible, infected, and removed individuals in Eqs. (4) to (6), where the infection/removal rate is a nonlinear function of the infected and susceptible populations. The derivation process of analytic expressions of the susceptible, infected, and removed populations is summarized in Fig. 2.

Figure 1 confirms that Eqs. (4) through (6) are correct and in agreement with computational simulations. We have compared the SIR model for epidemiology with the time derivatives of Eqs. (4) to (6) when the infection and removal rates are linear functions of the number of susceptible and infected individuals. However, we have found that the solutions of the differential equations for the SIR model are not given by Eqs. (4) to (6), as illustrated in Fig. 3.

We compare the infection dynamics under a nonlinear infection rate $\lambda_{n,s} = \dfrac{\beta ns}{1+\alpha n}$ with those from the standard linear form $\lambda_{n,s} = \beta ns$. Figure 4 shows that the nonlinear model exhibits slower growth and a lower peak, reflecting saturation effects at high infection levels. This comparison highlights the importance of nonlinear rates in accurately capturing infection dynamics. Several other examples illustrate the role of nonlinearity. The nonlinear infection rate $\lambda_{n,s} = \beta ns^{\alpha}$ $(\alpha \neq 1)$ implies limited contact when $\alpha < 1$ and high-density effects when $\alpha > 1$, reflecting heterogeneous contacts and deviations from homogeneous mixing. Behavioral changes in response to increasing infection levels can also be incorporated, for example with $\lambda_{n,s} = \beta nse^{-\alpha n}$, which reduce transmission as the number of infected individuals increases.

We have also examined the number of individuals in each compartment (S, I, or R) upon the arrival of infected individuals. The increase in the infected population stems from two sources: infections from the index case and external infections. This work derives the expression by considering the probability density function $f_{k,m,r}(t)$ of the time required to complete *m* infections and *r* arrivals with *k* infected individuals, written as Eq. (8). This relates to the probability $P_n(t)$ for the number *n* of infected individuals at time *t*, given in Eq. (9). Finally, Eq. (11) yields analytic expressions for the susceptible, infected, and removed populations. Figure 5 demonstrates that Eq. (11) is indeed correct. The most important aspect of Eqs. (4) through (6) and (11) is that they provide analytic expressions for each compartment in the infectious disease spread model, even though the infection and removal rates, denoted by $\lambda_{n,s}$ and $\mu_n$, respectively, are arbitrary functions of the number of susceptible and infected individuals.

It is easy to observe interacting subjects undergoing a birth-death process in which the birth and death rates are functions of the number of interacting compartments. For example, analytic expressions [Eqs. (4)-(6) and (11)] apply to consecutive reactions in chemistry, interspecific interactions in ecology, and other phenomena. The framework developed in this work to derive these expressions can be applied to study population dynamics, first-passage time(Park & Choi, 2024; Redner, 2001), queue waiting time (Gross et al., 2011; Kleinrock & Gail, 1996), and transport dynamics (Song et al., 2019; Spitzer, 2013). This framework provides a novel approach to analyzing quantitatively birth-death processes involving interacting subjects and feedback (i.e., autoregulation).

Our goal is to extend the method of deriving the analytic results to more complex and realistic birth-death processes involving finite death channels, more than three interacting subjects, and bulk births or deaths. These extensions have broad applications in diverse fields such as queueing theory, chemical kinetics, and epidemiology. Further research could clarify the time-evolution dynamics of the product number undergoing these birth-death processes. Ultimately, we could quantitatively understand and control these processes. Future work will explore a more realistic infectious disease spread model covering reinfection, emergence of infected individuals unlinked to the index case, multiple index cases at time zero, and unlimited population size.

## Acknowledgments

This work was supported by the National Research Foundation (NRF) grants funded by the Korean government (Ministry of Science and ICT) [Grant No. RS2021NR061970 (SJP), RS-2025-02663016(SJ)].

## Author Declarations

### Conflict of Interest

The authors have no conflicts of interest to declare.

### Data Availability

No data were used for the research described in the article.

## Supplementary Material

### Derivation of Eqs. (1) to (4)

This work focuses on an infectious disease transmission model in which the rates of infection and removal (recovery or death) depend on the number of infectious and susceptible individuals. In this model, a population exposed to an infection is divided into three groups: susceptible (S), infected (I), and removed (R, meaning recovered or deceased). The model satisfies the following three assumptions:

(i) There is one infected individual, the index case, at time zero.

(ii) The total population (the sum of the susceptible, infected, and removed individuals) remains constant $\mathrm{C} = S_0 + 1$, where $S_0$ denotes the number of susceptible individuals at time zero.

(iii) Removed individuals cannot become reinfected.

Let $\lambda_{n,s}$ denote the infection rate when the number of infectious individuals increases from $n$ to $n + 1$ and the number of susceptible individuals decreases from $s$ to $s - 1$. Also, let $\mu_n$ denote the removal rate when the number of infected individuals decreases from $n$ to $n - 1$. Under these assumptions, we express the probability $P_n(t)$ that the number of infected individuals is $n$ at time $t$ as a function of $\lambda_{n,s}$ and $\mu_n$. To obtain an analytic expression of $P_n(t)$, we introduce the probability rate function $f_{k,m}(t)$, which describes the distribution of times required for the $m$th infection to occur among $k$ infected individuals $(m = 1, 2, 3, \cdots; k = 2, 3, \cdots, m+1)$. By definition, we have $f_{k,m}(t) = 0$ when $k \le 1$ or $m \le 0$ because there must be at least two infected individuals at the time of infection.

First, we evaluate $f_{k,m}(t)$ under the initial condition $P_n(0) = \delta_{n1}$. The first infection can only occur if the index case does not recover or die. The number of infected individuals then becomes two immediately upon the first infection. Therefore, we can express $f_{2,1}(t)$ as $f_{2,1}(t) = \lambda_{1,S_0} e^{-(\lambda_{1,S_0} + \mu_1)t}$. When the infection occurs twice, the number of infected cases takes the value of two or three immediately upon the second infection, which makes it necessary to determine $f_{2,2}(t)$ and $f_{3,2}(t)$. For there to be two infected individuals at time *t* immediately after the second infection, the following process must have occurred: the first infection occurs at time $\tau_1$; the first infected individual is removed during the time interval $\tau_2$ without becoming infected; then the second infection occurs during the time interval $t-\tau_1-\tau_2\ (\geq 0)$. Mathematically, $f_{2,2}(t)$ is written as follows:

$$f_{2,2}(t)=\int_0^t f_{2,1}(\tau_1)\int_0^{t-\tau_1}\mu_2 e^{-(\mu_2+\lambda_{2,S_0-1})\tau_2}\lambda_{1,S_0-1}e^{-(\lambda_{1,S_0-1}+\mu_1)(t-\tau_1-\tau_2)}d\tau_2 d\tau_1 .$$

Suppose that the system size is three immediately after the second infection. This situation corresponds to the case in which the first infection occurs at time $\tau$ and the first infected individual survives until the second infection occurs during the time interval $t-\tau$. We thus obtain $f_{3,2}(t)=\int_0^t f_{2,1}(\tau)\lambda_{2,S_0-1}e^{-(\lambda_{2,S_0-1}+\mu_2)(t-\tau)}d\tau$.

Next, we carefully examine the case in which the third infection occurs. The number of infections must be two, three, or four immediately after the third infection. First, we focus on the case described by $f_{2,3}(t)$, where the number of infected individuals is two at the time of the third infection. There are two possibilities for $f_{2,3}(t)$: One possibility is that the infection count is two immediately after the second infection at time $\tau_1$, and the only product destroys itself during the time interval $\tau_2$, where the third infection occurs during $t-\tau_1-\tau_2$. The other possibility is that the infection count is three immediately after the second infection at time $\tau_1$, the three infected individuals destroy one during the time intervals $\tau_2$ and $\tau_3$, and the third infection occurs during $t-\tau_1-\tau_2-\tau_3$. A combination of these cases results in

$$\begin{aligned} f_{2,3}(t)&=\int_0^t f_{2,2}(\tau_1)\int_0^{t-\tau_1}\mu_2 e^{-(\mu_2+\lambda_{2,S_0-2})\tau_2}\lambda_{1,S_0-2}e^{-(\lambda_{1,S_0-2}+\mu_1)(t-\tau_1-\tau_2)}d\tau_2 d\tau_1 \\ &\quad+\int_0^t f_{3,2}(\tau_1)\int_0^{t-\tau_1}\mu_3 e^{-(\mu_3+\lambda_{3,S_0-2})\tau_2}\int_0^{t-\tau_1-\tau_2}\mu_2 e^{-(\mu_2+\lambda_{2,S_0-2})\tau_3}\lambda_{1,S_0-2}e^{-(\lambda_{1,S_0-2}+\mu_1)(t-\tau_1-\tau_2-\tau_3)}d\tau_3 d\tau_2 d\tau_1 \end{aligned} .$$

Second, we examine $f_{3,3}(t)$, where there are three infections immediately following the third infection. There are two possibilities for this: One possibility is that the number of infected individuals is two immediately after the second infection at time $\tau_1$ and the two infected individuals are not removed during the third infection time interval $t-\tau_1$. The second possibility is that the number of cases is three immediately after the second infection at time $\tau_1$. Then, one of the three infected individuals is removed at the rate $\mu_3$ without additional infection during the time interval $\tau_2$ while the other two remain infected during the time interval $t-\tau_1-\tau_2$. We thus have

$$f_{3,3}(t)=\int_0^t f_{2,2}(\tau)\lambda_{2,S_0-2}e^{-(\lambda_{2,S_0-2}+\mu_2)(t-\tau)d\tau}d\tau+\int_0^t f_{3,2}(\tau_1)\int_0^{t-\tau_1}\mu_3 e^{-(\mu_3+\lambda_{3,S_0-2})\tau_2}\lambda_{2,S_0-2}e^{-(\lambda_{2,S_0-2}+\mu_2)(t-\tau_1-\tau_2)}d\tau_2 d\tau_1$$

.

Finally, we focus on the case in which four individuals become infected at the time of the third infection. There is only one possibility for this: The number of infected individuals is three at time $\tau$ immediately after the second infection, and two of the infected organisms survive during the third infection time interval $t-\tau$. This leads to

$$f_{4,3}(t)=\int_0^t f_{3,2}(\tau)\lambda_{3,S_0-2}e^{-(\lambda_{3,S_0-2}+\mu_3)(t-\tau)}d\tau\ .$$

Similarly, we obtain $f_{k,4}(t)$ for $k$ = 2, 3, 4, and 5 as follows:

$$\begin{aligned}f_{2,4}(t)=&\int_0^t f_{2,3}(\tau_1)\int_0^{t-\tau_1}\mu_2 e^{-(\mu_2+\lambda_{2,S_0-3})\tau_2}\lambda_{1,S_0-3}e^{-(\lambda_{1,S_0-3}+\mu_1)(t-\tau_1-\tau_2)}d\tau_2 d\tau_1\\ &+\int_0^t f_{3,3}(\tau_1)\int_0^{t-\tau_1}\mu_3 e^{-(\mu_3+\lambda_{3,S_0-3})\tau_2}\int_0^{t-\tau_1-\tau_2}\mu_2 e^{-(\mu_2+\lambda_{2,S_0-3})\tau_3}\lambda_{1,S_0-3}e^{-(\lambda_{1,S_0-3}+\mu_1)(t-\tau_1-\tau_2-\tau_3)}d\tau_3 d\tau_2 d\tau_1\\ &+\int_0^t f_{4,3}(\tau_1)\int_0^{t-\tau_1}\mu_4 e^{-(\mu_4+\lambda_{4,S_0-3})\tau_2}\int_0^{t-\tau_1-\tau_2}\mu_3 e^{-(\mu_3+\lambda_{3,S_0-3})\tau_3}\int_0^{t-\tau_1-\tau_2-\tau_3}\mu_2 e^{-(\mu_2+\lambda_{2,S_0-3})\tau_4}\lambda_{1,S_0-3}e^{-(\lambda_{1,S_0-3}+\mu_1)(t-\tau_1-\tau_2-\tau_3-\tau_4)}d\tau_4 d\tau_3 d\tau_2 d\tau_1\end{aligned}$$

$$\begin{aligned}f_{3,4}(t)=&\int_0^t f_{2,3}(\tau)\lambda_{2,S_0-3}e^{-(\lambda_{2,S_0-3}+\mu_2)(t-\tau)}d\tau+\int_0^t f_{3,3}(\tau_1)\int_0^{t-\tau_1}\mu_3 e^{-(\mu_3+\lambda_{3,S_0-3})\tau_2}\lambda_{2,S_0-3}e^{-(\lambda_{2,S_0-3}+\mu_2)(t-\tau_1-\tau_2)}d\tau_2 d\tau_1\\ &+\int_0^t f_{4,3}(\tau_1)\int_0^{t-\tau_1}\mu_4 e^{-(\mu_4+\lambda_{4,S_0-3})\tau_2}\int_0^{t-\tau_1-\tau_2}\mu_3 e^{-(\mu_3+\lambda_{3,S_0-3})\tau_3}\lambda_{2,S_0-3}e^{-(\lambda_{2,S_0-3}+\mu_2)(t-\tau_1-\tau_2-\tau_3)}d\tau_3 d\tau_2 d\tau_1\end{aligned}$$

$$f_{4,4}(t)=\int_0^t f_{3,3}(\tau)\lambda_{3,S_0-3}e^{-(\lambda_{3,S_0-3}+\mu_3)(t-\tau)}d\tau+\int_0^t f_{4,3}(\tau_1)\int_0^{t-\tau_1}\mu_4 e^{-(\mu_4+\lambda_{4,S_0-3})\tau_2}\lambda_{3,S_0-3}e^{-(\lambda_{3,S_0-3}+\mu_3)(t-\tau_1-\tau_2)}d\tau_2 d\tau_1$$

$$f_{5,4}(t)=\int_0^t f_{4,3}(\tau)\lambda_{4,S_0-3}e^{-(\lambda_{4,S_0-3}+\mu_4)(t-\tau)}d\tau\ .$$

The probability density function $f_{k,m}(t)$ describing the distribution of times to complete the *m*th infection $(m=1,2,3,\cdots)$ among $k(=2,3,\cdots,m+1)$ infected individuals can be generalized via mathematical induction. The Laplace transform of $f_{k,m}(t)$ satisfies the following recursion relation:

$$
\begin{aligned}
\hat{f}_{k,m}(s) = {} & \hat{f}_{k-1,m-1}(s)\frac{\lambda_{k-1,S_0-k}}{s+\lambda_{k-1,S_0-k}+\mu_{k-1}} \\
& + \sum_{j=0}^{m-k}\hat{f}_{k+j,m-1}(s)\left(\prod_{l=0}^{j}\frac{\mu_{k+j-l}}{s+\mu_{k+j-l}+\lambda_{k+j-l,S_0-(m-1)}}\right)\frac{\lambda_{k-1,S_0-(m-1)}}{s+\mu_{k-1}+\lambda_{k-1,S_0-(m-1)}}
\end{aligned} \tag{A-1}
$$

with $\hat{f}(s)=\int_0^\infty f(t)e^{-st}dt$ and $\hat{f}_{2,1}(s)=\dfrac{\lambda_{1,S_0}}{s+\lambda_{1,S_0}+\mu_1}$. Upon inspection, we know that Eq. (A-1) holds for $m$ = 1 and 2. We now assume that Eq. (A-1) holds for $m = n$ and then prove that it holds for $m = n + 1$. From Eq. (A-1), to have $k$ infected individuals at the time of the $m$th infection, the number of infections must be equal to or greater than $k-1$ and equal to or less than $m-1$ at the time of the ($m-1$)th infection. Similarly, for $k$ infections to exist at the (m +1)th infection, the case count should be neither less than $k-1$ nor greater than $m$ at the $m$th infection. Using the Bromwich integral, we write the inverse Laplace transform of Eq. (A-1) in the form:

$$
\begin{aligned}
f_{k,m}(t) = {} & \int_0^t f_{k-1,m-1}(\tau)\lambda_{k-1,S_0-(m-1)}e^{-(\lambda_{k-1,S_0-(m-1)}+\mu_{k-1})(t-\tau)}d\tau \\
& + \sum_{j=0}^{m-k}\left\{\left(\prod_{l=0}^{j}\mu_{k+j-l}\right)\int_0^t f_{k+j,m-1}(\tau_1)\int_0^{t-\tau_1} g_{k,j,S_0}(\tau_2)\lambda_{k-1,S_0-(m-1)}e^{-(\lambda_{k-1,S_0-(m-1)}+\mu_{k-1})(t-\tau_1-\tau_2)}d\tau_2 d\tau_1\right\}
\end{aligned} \tag{A-2}
$$

with

$$
\begin{aligned}
g_{k,j,S_0}(\tau_2) &\equiv L^{-1}\left(\prod_{l=0}^{j}\left(s+\mu_{k+j-l}+\lambda_{k+j-l,S_0-(m-1)}\right)^{-1}\right)\\
&= \sum_{i=1}^{q}\sum_{z_1}\sum_{z_2}\cdots\sum_{z_{i-1}}\sum_{z_{i+1}}\cdots\sum_{z_q}\sum_{z_{q+1}}\cdots\sum_{z_{q+j-\sum_{i=1}^{q}y_i+2}} \left\{ \begin{array}{l} \dfrac{\prod\limits_{\substack{u=1\\u\neq i}}^{q}(-1)^{z_u}\dfrac{(y_u+z_u-1)!}{(y_u-1)!}(c_{k+j,S_0,u}-c_{k+j,S_0,i})^{-y_u-z_u}}{z_1!z_2!\cdots z_{i-1}!z_{i+1}!\cdots z_q!z_{q+1}!\cdots z_{q+j-\sum_{i=1}^{q}y_i+1}!} \\ \times\left(\prod\limits_{r=q+1}^{q+j-\sum_{i=1}^{q}y_i+1}(-1)^{z_r}(z_r!)(c_{k+j,S_0,r}-c_{k+j,S_0,i})^{-1-z_r}\right)\tau_2^{z_{q+j-\sum_{i=1}^{q}y_i+2}}e^{-c_{k+j,S_0,i}\tau_2}\end{array}\right\}\\
&\quad+\sum_{r=q+1}^{q+1+j-\sum_{i=1}^{q}y_i}\left\{\left(\prod_{i=1}^{q}(c_{k+j,S_0,i}-c_{k+j,S_0,r})^{-y_i}\right)\left(\prod_{\substack{v=q+1\\v\neq r}}^{q+1+j-\sum_{i=1}^{q}y_i}(c_{k+j,S_0,v}-c_{k+j,S_0,r})^{-1}\right)e^{-c_{k+j,S_0,r}\tau_2}\right\}
\end{aligned}
$$

where $z_w$'s (with $w=1,2,\cdots,q+j-\sum_{i=1}^{q}y_i+1$) are integers, ranging from 0 to $y_i-1$ (i.e., $0\le z_w\le y_i-1$) and satisfying $\sum_{\substack{w=1\\w\neq i}}^{q+j-\sum_{i=1}^{q}y_i+1} z_w = y_i-1$. Moreover, $c_{k+j,S_0,u}$ represents $\mu_{k+j-l}+\lambda_{k+j-l,S_0-(m-1)}$ (for $l=0,1,2,\cdots,j$), where $y_i$ denotes the multiplicity of $c_{k+j,S_0,i}$. For $\mu_{k+j-l}+\lambda_{k+j-l,S_0-(m-1)}$, $c_{k+j,S_0,i}$'s are ordered from high multiplicity and the largest value to low multiplicity and the smallest value. For example, we order $a_1=1$, $a_2=3$, $a_3=1$, $a_4=4$, $a_5=2$, $a_6=3$, and $a_7=5$ as $c_1=1=a_1=a_3$, $c_2=3=a_2=a_6$, $c_3=2=a_5$, $c_4=4=a_4$, and $c_5=5=a_7$. If infection and removal rates are all different form each other, we have $\lambda_i\neq\lambda_j$ and $\mu_i\neq\mu_j$ for any *i* and *j* ($\neq i$) and the first term in the right-hand side of the above equation for $g_{k,j,S_0}(\tau_2)$ vanishes. Further, it follows that

$$
\sum_{r=q+1}^{j+1}\left(\prod_{\substack{v=1\\v\neq r}}^{j+1}(c_{k+j,S_0,v}-c_{k+j,S_0,r})^{-1}\right)e^{-c_{k+j,S_0,r}\tau_2}.
$$

Next, we derive $P_n(t)$ from Eqs. (1) and (2). First, we consider the case of one infection at time *t* in two situations. In the first situation, the initial condition $P_n(0)=\delta_{n1}$ means that the first infection does not occur until time $t$ without recovery or death. In the second situation, all infected individuals are removed except for one case existing at time $t$. The probability that the first infection would not occur until time $t$ without removal of the index case reads $e^{-(\lambda_{1,S_0}+\mu_1)t}$. Then, we consider the case of one infected individual who has been infected at least once. This work assumes that $k$ infected individuals exist when the $m$th infection occurs at time $\tau_1$ and that $k - 1$ cases of infection are removed during the interval $\tau_2$.

For a single infection to exist at time $t$, there must be no infections prior to $t$; this case is described by

$$\int_0^t f_{k,m}(\tau_1)\int_0^{t-\tau_1}\left(\prod_{l=0}^{k-2}\mu_{k-l}\right)h_{k,k-2,S_0}(\tau_2)e^{-(\lambda_{1,S_0-m}+\mu_1)}d\tau_2 d\tau_1$$

where $h_{k,j,S_0}(t)$ denotes the inverse Laplace transform of

$\hat{h}_{k,j,S_0}(s)=\int_0^\infty h_{k,j,S_0}(t)e^{-st}dt=\prod_{l=0}^{j}(s+\mu_{k-l}+\lambda_{k-l,S_0-m})^{-1}$. As $m$ and $k$ are positive integers satisfying $m\geq 1$ and $2\leq k\leq m+1$, we have

$P_1(t)=e^{-(\lambda_{1,S_0}+\mu_1)t}+\sum_{m=1}^{S_0}\sum_{k=2}^{m+1}\int_0^t f_{k,m}(\tau_1)\int_0^{t-\tau_1}\left(\prod_{l=0}^{k-2}\mu_{k-l}\right)h_{k,k-2,S_0}(\tau_2)e^{-(\lambda_{1,S_0-m}+\mu_1)(t-\tau_1-\tau_2)}d\tau_2 d\tau_1$. The first term, $e^{-(\lambda_{1,S_0}+\mu_1)t}$, of $P_1(t)$ equals $\frac{f_{2,1}(t)}{\lambda_{1,S_0}}$, and the last term,

$\sum_{m=1}^{S_0}\sum_{k=2}^{m+1}\int_0^t f_{k,m}(\tau_1)\int_0^{t-\tau_1}\left(\prod_{l=0}^{k-2}\mu_{k-l}\right)h_{k,j,S_0}(\tau_2)e^{-(\lambda_{1,S_0-m}+\mu_1)(t-\tau_1-\tau_2)}d\tau_2 d\tau_1$, reduces to $\sum_{m=2}^{S_0}\frac{f_{2,m}(t)}{\lambda_{1,S_0-(m-1)}}$ because substituting $k = 2$ in Eq. (A-2) results in

$$f_{2,m}(t)=\sum_{j=0}^{m-2}\left\{\left(\prod_{l=0}^{j}\mu_{2+j-l}\right)\int_0^t f_{2+j,m-1}(\tau_1)\int_0^{t-\tau_1}g_{2,j,S_0}(\tau_2)\lambda_{1,S_0-(m-1)}e^{-(\lambda_{1,S_0-(m-1)}+\mu_1)(t-\tau_1-\tau_2)}d\tau_2 d\tau_1\right\}.$$

Consequently, $P_1(t)$ can be simplified considerably into the expression:

$P_1(t)=\sum_{m=1}^{S_0}\frac{f_{2,m}(t)}{\lambda_{1,S_0-(m-1)}}$. As for $P_2(t)$, there should be at least one infection by time $t$, and two possibilities contribute to $P_2(t)$: One is that there are two infections immediately after the $m$th birth at time $\tau_1$, and the two infected individuals survive without infection or removal during $t-\tau_1$, resulting in $\sum_{m=2}^{S_0}\int_0^t f_{2,m}(\tau)e^{-(\lambda_{2,S_0-m}+\mu_2)(t-\tau)}d\tau$. The other possibility is that there exist $k$ infected individuals at time $\tau_1$ immediately after the $m$th birth, $k-2$ of whom recover or die during $\tau_2$, leaving two infected individuals who survive without additional infection or removal during $t-\tau_1-\tau_2$. This possibility yields

$$\sum_{m=2}^{S_0}\sum_{k=3}^{m+1}\int_0^t f_{k,m}(\tau_1)\int_0^{t-\tau_1} h_{k,k-3,S_0}(\tau_2)e^{-(\lambda_{2,S_0-m}+\mu_2)(t-\tau_1-\tau_2)}d\tau_2 d\tau_1 .$$

Thus, $P_2(t)$ is expressed as follows:

$$P_2(t)=\sum_{m=1}^{S_0}\sum_{k=2}^{m+1}\int_0^t f_{k,m}(\tau_1)\left(\delta_{2,k}e^{-(\lambda_{2,S_0-m}+\mu_2)(t-\tau_1)}+(1-\delta_{2,k})\int_0^{t-\tau_1} h_{k,k-3,S_0}(\tau_2)e^{-(\lambda_{2,S_0-m}+\mu_2)(t-\tau_1-\tau_2)}d\tau_2\right)d\tau_1 ,$$

where $\delta_{a,b}$ denotes the Kronecker delta. Similarly, $P_3(t)$ is given by

$$P_3(t)=\sum_{m=2}^{S_0}\sum_{k=3}^{m+1}\int_0^t f_{k,m}(\tau_1)\left(\delta_{3,k}e^{-(\lambda_{3,S_0-m}+\mu_3)(t-\tau_1)}+(1-\delta_{3,k})\int_0^{t-\tau_1} h_{k,k-4,S_0}(\tau_2)e^{-(\lambda_{3,S_0-m}+\mu_3)(t-\tau_1-\tau_2)}d\tau_2\right)d\tau_1 .$$

Furthermore, $P_2(t)$ and $P_3(t)$ also simplify to the following expressions:

$P_2(t)=\sum_{m=2}^{S_0}\frac{f_{3,m}(t)}{\lambda_{2,S_0-(m-1)}}$ and $P_3(t)=\sum_{m=3}^{S_0}\frac{f_{4,m}(t)}{\lambda_{3,S_0-(m-1)}}$. Deriving $P_n(t)$ for $n=2,\cdots,S_0$ is similar to deriving $P_1(t)$, where $P_0(t)$ and $P_{S_0+1}(t)$ must be noted because infection and removal rates cannot be defined when there are no infected or susceptible individuals. We categorize the case of zero infected individuals at time $t$ as two events: One event is that the infection does not occur until time $t$; the other is that all infected individuals are removed, resulting in no infected individuals at time $t$. The index case must either recover or die for there to be no infected individuals without the infection occurring. The probability that the index case will recover or die by time $t$ is $\int_0^t \mu_1 e^{-(\mu_1+\lambda_{1,S_0})\tau}d\tau$.

We consider the case of an initially uninfected population in the presence of an infection that occurs at least once. If $k$ infected individuals exist when the $m$th infection occurs at time $\tau_1$ and no infected individuals to exist at time $\tau$, then the *k* infected cases must have been removed during the interval $\tau-\tau_1$. This case is described by $\int_0^t\int_0^\tau f_{k,m}(\tau_1)v_{k,m,S_0}(\tau-\tau_1)d\tau_1 d\tau$, where $v_{k,m,S_0}(t)$ denotes the inverse Laplace transform of $\hat{v}_{k,m,S_0}(s)=\int_0^\infty v_{k,m,S_0}(t)e^{-st}dt=\prod_{l=1}^{k}\frac{\mu_l}{s+\mu_l+\lambda_{l,S_0-m}}$. Then, we obtain

$$P_0(t)=\int_0^t\left(\mu_1 e^{-(\mu_1+\lambda_{1,S_0})\tau}+\sum_{m=1}^{\infty}\sum_{k=2}^{m+1}\int_0^\tau f_{k,m}(\tau_1)v_{k,m,S_0}(\tau-\tau_1)d\tau_1\right)d\tau .$$

Compared with $P_1(t)$, we can rewrite $P_0(t)$ in the form $P_0(t)=\int_0^t\mu_1 P_1(\tau)d\tau$. For $P_{S_0+1}(t)$, we consider the possible case that the number of infected individuals is $S_0+1$. The infection must occur $S_0$ times for the infection count to be $S_0+1$. There is one possibility for the infection count to be $S_0+1$: The infected population is $S_0+1$ at time $\tau$, and recovery or death must not occur during the time interval $t-\tau$. This possibility results in $P_{S_0+1}(t)=\int_0^t f_{S_0+1,S_0}(\tau)e^{-\mu_{S_0+1}(t-\tau)}d\tau$.

Summarizing the result of $P_n(t)$ for $n=0,1,2,\cdots,S_0+1$, we complete the proof of Eq. (1). The sum of $P_n(t)$ over *n* is rearranged as follows:

$$\begin{aligned}\sum_{n=0}^{S_0+1}P_n(t)&=1-\int_0^t f_{2,1}(\tau)d\tau+\left(\int_0^t f_{2,1}(\tau)d\tau-\sum_{i=2}^{3}\int_0^t f_{i,2}(\tau)d\tau\right)+\left(\sum_{i=2}^{3}\int_0^t f_{i,2}(\tau)d\tau-\sum_{i=2}^{4}\int_0^t f_{i,3}(\tau)d\tau\right)\\&\quad+\cdots+\left(\sum_{i=2}^{S_0}\int_0^t f_{i,S_0-1}(\tau)d\tau-\sum_{i=2}^{S_0+1}\int_0^t f_{i,S_0}(\tau)d\tau\right)+\sum_{i=2}^{S_0+1}\int_0^t f_{i,S_0}(\tau)d\tau=1\end{aligned},$$

which completes the proof of $\sum_{n=0}^{S_0+1}P_n(t)=1$.

We derive a differential equation for $P_n(t)$ using a Laplace transform. Let the Laplace transform of a function *f*(*t*) be denoted by $\hat{f}(s)=\int_0^\infty f(t)e^{-st}dt$. Examining the Laplace transform of $P_1(t)$, we find $s\hat{P}_1(s)-1=-\sum_{m=1}^{S_0}\hat{f}_{2,m}(s)-\mu_1\hat{P}_1(s)+\mu_2\hat{P}_2(s)$ or

$s\hat{P}_1(s) - P_1(0) = -\sum_{m=1}^{S_0} \hat{f}_{2,m}(s) - \mu_1 \hat{P}_1(s) + \mu_2 \hat{P}_2(s)$ due to the initial condition $P_n(0) = \delta_{n1}$. As for $\hat{P}_2(s)$, we determine that $s\hat{P}_2(s) = -\sum_{m=2}^{S_0} \hat{f}_{3,m}(s) - \mu_2 \hat{P}_2(s) + \sum_{m=1}^{S_0} \hat{f}_{2,m}(s) + \mu_3 \hat{P}_3(s)$. Similarly, it is straightforward to show that $s\hat{P}_n(s) = -\sum_{m=n}^{S_0} \hat{f}_{n+1,m}(s) - \mu_n \hat{P}_n(s) + \sum_{m=n-1}^{S_0} \hat{f}_{n,m}(s) + \mu_{n+1} \hat{P}_{n+1}(s)$ for $n = 2, 3, \cdots, S_0$. We can easily derive $s\hat{P}_0(s) = \mu_1 \hat{P}_1(s)$ and $s\hat{P}_{S_0+1}(s) = \hat{f}_{S_0+1,S_0}(s)\left(1 - \frac{\mu_{S_0+1}}{s + \mu_{S_0+1}}\right)$ because $P_0(t) = \int_0^t \mu_1 P_1(\tau) d\tau$ and $P_{S_0+1}(t) = \int_0^t f_{S_0+1,S_0}(\tau) e^{-\mu_{S_0+1}(t-\tau)} d\tau$, respectively. The inverse Laplace transform of $s\hat{P}_n(s)$ for $n = 0, 1, 2, \cdots, S_0 + 1$ gives Eq. (3).

**Derivation of Eqs. (8) to (11)**

This work provides a detailed discussion of an extended infectious disease spread model. In the model, newly infected individuals emerge at a specific time and can infect susceptible individuals. They are not infected by those who were infected by the identified index case at time zero. The model incorporates the newly infected individuals into the infectious disease transmission model described by Eq. (1). This work assumes that the number of infected individuals introduced to the system is $v$ and that the times when new cases are introduced are denoted by $a_1, a_2, \cdots, a_v$. The total population becomes $S_0 + v + 1$ when the time exceeds $a_v$. Hence, infected organisms fall into two categories: those transmitted from a contagious individual and those infected due to an introduced case occurring at some time among $a_1, a_2, \cdots, a_v$. We then introduce the probability density function $f_{k,m,r}(t)$ of the time ($a_r \le t < a_{r+1}$) to complete $m$ infections, where $k$ is the number of infected individuals immediately after the $m$th infection ($k \ge 1, m \ge 0, r \ge 0$). The following paragraph discusses $f_{k,m,r}(t)$.

Now, we evaluate $f_{k,m,r}(t)$ under the initial condition $P_n(0) = \delta_{n1}$. We consider $f_{k,m,r}(t)$ when $m + r = 1$. The values of $m$ and $r$ are both greater than or equal to zero, and there are two possibilities: $m = 0$ and $r = 1$, or $m = 1$ and $r = 0$. When $m = 0$ and $r = 1$, if the index case recovers or dies during the time interval $a_1$, the number of infected individuals is one when the first new infected individual appears. If the index case is not removed (recovers or dies), the infection count is two at time $a_1$. Accordingly, $f_{1,0,1}(t)$ and $f_{2,0,1}(t)$ are expressed as $f_{1,0,1}(t) = \delta(t - a_1) \int_0^{a_1} \mu_1 e^{-(\mu_1 + \lambda_{1,S_0})\tau} d\tau$ and $f_{2,0,1}(t) = \delta(t - a_1) e^{-(\mu_1 + \lambda_{1,S_0}) a_1}$, respectively. For $m = 1$ and $r = 0$, the case count is two immediately after the first infection. The index case must survive and infect one susceptible individual before time $t$; an imported case must not appear before time $t$. Thus, $f_{2,1,0}(t)$ is expressed as $f_{2,1,0}(t) = \lambda_{1,S_0} e^{-(\lambda_{1,S_0} + \mu_1)t} \theta(a_1 - t)$ because $1 - \int_0^t \delta(\tau - a_1) d\tau = \theta(a_1 - t)$, where $\theta(x)$ is the Heaviside step function. One can easily demonstrate that $\int_0^\infty f_{1,0,1}(t) + f_{2,0,1}(t) + f_{2,1,0}(t) dt = 1$. Second, we consider the probability density function $f_{k,m,r}(t)$ of the time to complete *m* infections and *r* introduced cases for an infected count of *k* ($\leq 3$) and *m* + *r* = 2 at the *m*th infection or the *r*th arrival. When *m* + *r* = 2, there are three possible pairs (*m*, *r*) = (0, 2), (1, 1), and (2, 0). For the pair (0, 2), the number of infected individuals can be one, two, or three at arrival. If there is one case at the second arrival, there exist two possibilities contributing to $f_{1,0,2}(t)$. One is that the infected count is one immediately after the first arrival time $\tau_1$, where the only infected individual is removed during the time interval $a_2 - \tau_1$ and the second introduced case occurs at time $a_2$. The other possibility is that the infected population count is two at the first arrival time $\tau_1$, where two infected individuals are removed during the time intervals $\tau_2$ and $\tau_3$, and the second introduced case occur at time $a_2$. This possibility results in

$$\begin{aligned} f_{1,0,2}(t) &= \int_0^{a_2} f_{1,0,1}(\tau_1) \int_0^{a_2-\tau_1} \mu_1 e^{-(\mu_1+\lambda_{1,S_0})x} dx d\tau_1 \delta(t-a_2) \\ &+ \int_0^{a_2} f_{2,0,1}(\tau_1) \int_0^{a_2-\tau_1} \mu_2 e^{-(\mu_2+\lambda_{2,S_0})\tau_2} \int_0^{a_2-\tau_1-\tau_2} \mu_1 e^{-(\mu_1+\lambda_{1,S_0})\tau_3} d\tau_3 d\tau_2 \, d\tau_1 \delta(t-a_2) \end{aligned}.$$

This work examines $f_{2,0,2}(t)$, where the case count is two at the second arrival. There are two possibilities here: One possibility is that the number of infected individuals is one at the first arrival time $\tau_1$ and that the infected individual survives without additional infection until the second arrival. The other possibility is that two individuals are infected at the time of the first arrival, $\tau_1$. Then, one of the two infected individuals either recovers or dies without becoming infected again during $\tau_2$, while the other survives until the second arrival. We thus have

$$\begin{aligned} f_{2,0,2}(t) &= \int_0^{a_2} f_{1,0,1}(\tau_1) e^{-(\lambda_{1,S_0}+\mu_1)(t-\tau_1)} d\tau_1 \delta(t-a_2) \\ &+ \int_0^{a_2} f_{2,0,1}(\tau_1) \int_0^{a_2-\tau_1} \mu_2 e^{-(\mu_2+\lambda_{2,S_0})\tau_2} e^{-(\mu_1+\lambda_{1,S_0})(a_2-\tau_1-\tau_2)} d\tau_2 d\tau_1 \delta(t-a_2) \end{aligned}.$$

The case in which the number of infected individuals is three at the second arrival is also evaluated. Only one possibility exists for $f_{3,0,2}(t)$. There are two infected individuals at the first arrival at time $\tau_1$, they survive without additional infection during the time interval $a_2 - \tau_1$, and the second arrival occurs at time *t*. This yields $f_{3,0,2}(t) = \int_0^{a_2} f_{2,0,1}(\tau_1) e^{-(\lambda_{2,S_0}+\mu_2)(t-\tau_1)} d\tau_1 \delta(t-a_2)$.

Next, we examine the pair (1, 1), where the infected population may consist of one, two, or three individuals immediately after infection or arrival. We derive the probability density function $f_{k,1,1}(t)$ of the time it takes to complete the first infections and the first case of arrival with *k* infected individuals at that moment. There is one possibility for one infection count ($k = 1$) for $f_{1,1,1}(t)$. At the first infection instance at time $\tau_1$, two individuals are infected; two carriers are removed during the time interval $a_1 - \tau_1$, and the first influx occurs at time *t*. We thus have

$$f_{1,1,1}(t) = \int_0^{a_1} f_{2,1,0}(\tau_1) \int_0^{a_1-\tau_1} \mu_2 e^{-(\mu_2+\lambda_{2,S_0-1})\tau_2} \int_0^{a_1-\tau_1-\tau_2} \mu_1 e^{-(\mu_1+\lambda_{1,S_0-1})\tau_3} d\tau_3 d\tau_2 d\tau_1 \delta(t-a_1) .$$

Second, we assess $f_{2,1,1}(t)$ more closely. There are three possibilities that contribute to $f_{2,1,1}(t)$. The first scenario is that there is one case at the first arrival time $\tau_1$, where the first infection occurs during the time interval $t-\tau_1$, and the second arrival must occur during the time *t*. The second scenario is that there are two infections at the first arrival time $\tau_1$, where one of the two cases is removed without a new infection during the time $\tau_2$, the first infection occurs during the time $t-\tau_1-\tau_2$, and the second arrival case must not occur before time *t*. The final scenario is that the number of infected individuals is two immediately after the first infection at time $\tau_1$, where one of the two cases recovers or dies during the time interval $\tau_2$, infection or removal does not occur during the time interval $a_1-\tau_1-\tau_2$, and the first arrival occurs at time *t*. Accordingly, $f_{2,1,1}(t)$ is expressed as

$$\begin{aligned} f_{2,1,1}(t) = &\int_0^{t} f_{1,0,1}(\tau_1) \lambda_{1,S_0} e^{-(\lambda_{1,S_0}+\mu_1)(t-\tau_1)} d\tau_1 \theta(a_2-t) \\ &+ \int_0^{t} f_{2,0,1}(\tau_1) \int_0^{t-\tau_1} \mu_2 e^{-(\mu_2+\lambda_{2,S_0})\tau_2} \lambda_{1,S_0} e^{-(\lambda_{1,S_0}+\mu_1)(t-\tau_1-\tau_2)} d\tau_2 d\tau_1 \theta(a_2-t) . \\ &+ \int_0^{a_1} f_{2,1,0}(\tau_1) \int_0^{a_1-\tau_1} \mu_2 e^{-(\mu_2+\lambda_{2,S_0})\tau_2} e^{-(\mu_1+\lambda_{1,S_0-1})(a_1-\tau_1-\tau_2)} d\tau_2 d\tau_1 \delta(t-a_1) \end{aligned}$$

Regarding $f_{3,1,1}(t)$, there are two possibilities for the infection count to be three immediately upon the first infection and the first arrival at time *t*. One possibility is that the number of cases is two immediately upon the first arrival at time $\tau_1$, where the infection occurs without removal during the time interval $t-\tau_1$, and the second arrival does not occur before time $t$. The other possibility is that the infection count is two at the time $\tau_1$ of the first infection, where the infection occurs without removal during the time interval $a_1-\tau_1$, and the first arrival occurs at time $a_1$. This gives

$$f_{3,1,1}(t)=\int_0^t f_{2,0,1}(\tau_1)\lambda_{2,S_0}e^{-(\lambda_{2,S_0}+\mu_2)(t-\tau_1)}d\tau_1\theta(a_2-t)+\int_0^{a_1} f_{2,1,0}(\tau_1)e^{-(\lambda_{2,S_0-1}+\mu_2)(a_1-\tau_1)}d\tau_1\delta(t-a_1).$$

Moreover, we have $f_{1,2,0}(t)=0$ because the number of infected individuals is at least two at the infection instance. When $m$ = 2 and $r$ = 0, the case count can be two or three. We derive the probability density function $f_{2,2,0}(t)$ of waiting times until two infections occur with two infected individuals at the second infection. This ensures that the arrival of infected individuals does not occur before two infections are completed. Then, $f_{2,2,0}(t)$ is given by the product of $f_{2,2}(t)$ and $\theta(a_1-t)$:

$$f_{2,2,0}(t)=\int_0^t f_{2,1,0}(\tau_1)\int_0^{t-\tau_1}\mu_2e^{-(\lambda_{2,S_0-1}+\mu_2)\tau_2}\lambda_{1,S_0-1}e^{-(\lambda_{1,S_0-1}+\mu_1)(t-\tau_1-\tau_2)}d\tau_2d\tau_1\theta(a_1-t).$$

Similarly, $f_{3,2,0}(t)$ equals $f_{3,2}(t)$ multiplied by $\theta(a_1-t)$ and reads

$$f_{3,2,0}(t)=\int_0^t f_{2,1,0}(\tau_1)\lambda_{2,S_0-1}e^{-(\lambda_{2,S_0-1}+\mu_2)(t-\tau_1)}d\tau_1\theta(a_1-t).$$

Third, we examine $f_{k,m,r}(t)$ for $m$ + $r$ = 3, where $k=1,2,3,4$. Similarly, we derive $f_{k,m,r}(t)$ for $m$ + $r$ = 3:

$$\begin{aligned}
f_{1,0,3}(t)=&\int_0^{a_3} f_{1,0,2}(\tau_1)\int_0^{a_3-\tau_1}\mu_1e^{-(\mu_1+\lambda_{1,S_0})\tau_2}d\tau_2d\tau_1\delta(t-a_3)\\
&+\int_0^{a_3} f_{2,0,2}(\tau_1)\int_0^{a_3-\tau_1}\mu_2e^{-(\mu_2+\lambda_{2,S_0})\tau_2}\int_0^{a_3-\tau_1-\tau_2}\mu_1e^{-(\mu_1+\lambda_{1,S_0})\tau_3}d\tau_3d\tau_2d\tau_1\delta(t-a_3)\\
&+\int_0^{a_3} f_{3,0,2}(\tau_1)\int_0^{a_3-\tau_1}\mu_3e^{-(\mu_3+\lambda_{3,S_0})\tau_2}\int_0^{a_3-\tau_1-\tau_2}\mu_2e^{-(\mu_2+\lambda_{2,S_0})\tau_3}\int_0^{a_3-\tau_1-\tau_2-\tau_3}\mu_1e^{-(\mu_1+\lambda_{1,S_0})\tau_4}d\tau_4d\tau_3d\tau_2d\tau_1
\end{aligned}$$

$$
\begin{aligned}
f_{2,0,3}(t) &= \int_0^{a_3} f_{1,0,2}(\tau_1) e^{-(\mu_1+\lambda_{1,S_0})(t-\tau_1)} d\tau_1 \delta(t-a_3) \\
&+ \int_0^{a_3} f_{2,0,2}(\tau_1) \int_0^{t-\tau_1} \mu_2 e^{-(\mu_2+\lambda_{2,S_0})\tau_2} e^{-(\mu_1+\lambda_{1,S_0})(t-\tau_1-\tau_2)} d\tau_2 d\tau_1 \delta(t-a_3) \\
&+ \int_0^{a_3} f_{3,0,2}(\tau_1) \int_0^{t-\tau_1} \mu_3 e^{-(\mu_3+\lambda_{3,S_0})\tau_2} \int_0^{t-\tau_1-\tau_2} \mu_2 e^{-(\mu_2+\lambda_{2,S_0})\tau_3} e^{-(\mu_1+\lambda_{1,S_0})(t-\tau_1-\tau_2-\tau_3)} d\tau_3 d\tau_2 d\tau_1 \delta(t-a_3)
\end{aligned}
$$

$$
\begin{aligned}
f_{3,0,3}(t) &= \int_0^{a_3} f_{2,0,2}(\tau_1) e^{-(\mu_2+\lambda_{2,S_0})(t-\tau_1)} d\tau_1 \delta(t-a_3) \\
&+ \int_0^{a_3} f_{3,0,2}(\tau_1) \int_0^{t-\tau_1} \mu_3 e^{-(\mu_3+\lambda_{3,S_0})\tau_2} e^{-(\mu_2+\lambda_{2,S_0})(t-\tau_1-\tau_2)} d\tau_2 d\tau_1 \delta(t-a_3)
\end{aligned}
$$

$$
f_{4,0,3}(t) = \int_0^{a_3} f_{3,0,2}(\tau_1) e^{-(\mu_3+\lambda_{3,S_0})(t-\tau_1)} d\tau_1 \delta(t-a_3)
$$

$$
\begin{aligned}
f_{1,1,2}(t) &= \int_0^{a_2} f_{1,1,1}(\tau_1) \int_0^{a_2-\tau_1} \mu_1 e^{-(\mu_1+\lambda_{1,S_0-1})\tau_2} d\tau_2 d\tau_1 \delta(t-a_2) \\
&+ \int_0^{a_2} f_{2,1,1}(\tau_1) \int_0^{a_2-\tau_1} \mu_2 e^{-(\mu_2+\lambda_{2,S_0-1})\tau_2} \int_0^{a_2-\tau_1-\tau_2} \mu_1 e^{-(\mu_1+\lambda_{1,S_0-1})\tau_3} d\tau_3 d\tau_2 d\tau_1 \delta(t-a_2) \\
&+ \int_0^{a_2} f_{3,1,1}(\tau_1) \int_0^{a_2-\tau_1} \mu_3 e^{-(\mu_3+\lambda_{3,S_0-1})\tau_2} \int_0^{a_2-\tau_1-\tau_2} \mu_2 e^{-(\mu_2+\lambda_{2,S_0-1})\tau_3} \int_0^{a_2-\tau_1-\tau_2-\tau_3} \mu_1 e^{-(\mu_1+\lambda_{1,S_0-1})\tau_4} d\tau_4 d\tau_3 d\tau_2 d\tau_1 \delta(t-a_2)
\end{aligned}
$$

$$
\begin{aligned}
f_{2,1,2}(t) &= \int_0^{t} f_{1,0,2}(\tau_1) \lambda_{1,S_0} e^{-(\lambda_{1,S_0}+\mu_1)(t-\tau_1)} d\tau_1 \theta(a_3-t) \\
&+ \int_0^{t} f_{2,0,2}(\tau_1) \int_0^{t-\tau_1} \mu_2 e^{-(\lambda_{2,S_0}+\mu_2)\tau_2} \lambda_{1,S_0} e^{-(\lambda_{1,S_0}+\mu_1)(t-\tau_1-\tau_2)} d\tau_2 d\tau_1 \theta(a_3-t) \\
&+ \int_0^{t} f_{3,0,2}(\tau_1) \int_0^{t-\tau_1} \mu_2 e^{-(\lambda_{2,S_0}+\mu_2)\tau_2} \int_0^{t-\tau_1-\tau_2} \mu_2 e^{-(\lambda_{2,S_0}+\mu_2)\tau_3} \lambda_{1,S_0} e^{-(\lambda_{1,S_0}+\mu_1)(t-\tau_1-\tau_2-\tau_3)} d\tau_3 d\tau_2 d\tau_1 \theta(a_3-t) \\
&+ \int_0^{a_2} f_{1,1,1}(\tau_1) e^{-(\lambda_{1,S_0-1}+\mu_1)(a_2-\tau_1)} d\tau_1 \delta(t-a_2) \\
&+ \int_0^{a_2} f_{2,1,1}(\tau_1) \int_0^{a_2-\tau_1} \mu_2 e^{-(\lambda_{2,S_0-1}+\mu_2)\tau_2} e^{-(\lambda_{1,S_0-1}+\mu_1)(t-\tau_1-\tau_2)} d\tau_2 d\tau_1 \delta(t-a_2) \\
&+ \int_0^{a_2} f_{3,1,1}(\tau_1) \int_0^{a_2-\tau_1} \mu_3 e^{-(\lambda_{3,S_0-1}+\mu_3)\tau_2} \int_0^{a_2-\tau_1-\tau_2} \mu_2 e^{-(\lambda_{2,S_0-1}+\mu_2)\tau_3} e^{-(\lambda_{1,S_0}+\mu_1)(a_2-\tau_1-\tau_2-\tau_3)} d\tau_3 d\tau_2 d\tau_1 \delta(t-a_2)
\end{aligned}
$$

$$
\begin{aligned}
f_{3,1,2}(t) &= \int_0^{t} f_{2,0,2}(\tau_1) \lambda_{2,S_0} e^{-(\lambda_{2,S_0}+\mu_2)(t-\tau_1)} d\tau_1 \theta(a_3-t) \\
&+ \int_0^{t} f_{3,0,2}(\tau_1) \int_0^{t-\tau_1} \mu_3 e^{-(\lambda_{3,S_0}+\mu_3)\tau_2} \lambda_{2,S_0} e^{-(\lambda_{2,S_0}+\mu_2)(t-\tau_1-\tau_2)} d\tau_2 d\tau_1 \theta(a_3-t) \\
&+ \int_0^{a_2} f_{2,1,1}(\tau_1) e^{-(\lambda_{2,S_0-1}+\mu_2)(a_2-\tau_1)} d\tau_1 \delta(t-a_2) \\
&+ \int_0^{a_2} f_{3,1,1}(\tau_1) \int_0^{a_2-\tau_1} \mu_3 e^{-(\lambda_{3,S_0-1}+\mu_3)\tau_2} e^{-(\lambda_{2,S_0-1}+\mu_2)(a_2-\tau_1-\tau_2)} d\tau_2 d\tau_1 \delta(t-a_2)
\end{aligned}
$$

$$
f_{4,1,2}(t) = \int_0^{a_2} f_{3,1,1}(\tau_1) e^{-(\lambda_{3,S_0-1}+\mu_3)(t-\tau_1)} d\tau_1 \delta(t-a_2)
$$

$$
\begin{aligned}
f_{1,2,1}(t) &= \int_0^{a_1} f_{2,2,0}(\tau_1) \int_0^{a_1-\tau_1} \mu_2 e^{-(\mu_2+\lambda_{2,S_0-2})\tau_2} \int_0^{a_1-\tau_1-\tau_2} \mu_1 e^{-(\mu_1+\lambda_{1,S_0-2})\tau_3} d\tau_3 d\tau_2 d\tau_1 \delta(t-a_1) \\
&+ \int_0^{a_1} f_{3,2,0}(\tau_1) \int_0^{a_1-\tau_1} \mu_3 e^{-(\mu_3+\lambda_{3,S_0-2})\tau_2} \int_0^{a_1-\tau_1-\tau_2} \mu_2 e^{-(\mu_2+\lambda_{2,S_0-2})\tau_3} \int_0^{a_1-\tau_1-\tau_2-\tau_3} \mu_1 e^{-(\mu_1+\lambda_{1,S_0-2})\tau_4} d\tau_4 d\tau_3 d\tau_2 d\tau_1 \delta(t-a_1)
\end{aligned}
$$

$$
\begin{aligned}
f_{2,2,1}(t) &= \int_0^{t} f_{1,1,1}(\tau_1) \lambda_{1,S_0-1} e^{-(\lambda_{1,S_0-1}+\mu_1)(t-\tau_1)} d\tau_1 \theta(a_2-t) \\
&+ \int_0^{t} f_{2,1,1}(\tau_1) \int_0^{t-\tau_1} \mu_2 e^{-(\lambda_{2,S_0-1}+\mu_2)\tau_2} \lambda_{1,S_0-1} e^{-(\lambda_{1,S_0-1}+\mu_1)(t-\tau_1-\tau_2)} d\tau_2 d\tau_1 \theta(a_2-t) \\
&+ \int_0^{t} f_{3,1,1}(\tau_1) \int_0^{t-\tau_1} \mu_3 e^{-(\lambda_{3,S_0-1}+\mu_2)\tau_2} \int_0^{t-\tau_1-\tau_2} \mu_2 e^{-(\lambda_{2,S_0-1}+\mu_2)\tau_3} \lambda_{1,S_0-1} e^{-(\lambda_{1,S_0-1}+\mu_1)(t-\tau_1-\tau_2-\tau_3)} d\tau_3 d\tau_2 d\tau_1 \theta(a_2-t) \\
&+ \int_0^{a_1} f_{2,2,0}(\tau_1) \int_0^{a_1-\tau_1} \mu_2 e^{-(\lambda_{2,S_0-2}+\mu_2)\tau_2} e^{-(\lambda_{1,S_0-2}+\mu_1)(a_1-\tau_1-\tau_2)} d\tau_2 d\tau_1 \, \delta(t-a_1) \\
&+ \int_0^{a_1} f_{3,2,0}(\tau_1) \int_0^{a_1-\tau_1} \mu_3 e^{-(\lambda_{3,S_0-2}+\mu_3)\tau_2} \int_0^{a_1-\tau_1-\tau_2} \mu_2 e^{-(\lambda_{2,S_0-2}+\mu_2)\tau_3} e^{-(\lambda_{1,S_0-2}+\mu_1)(a_1-\tau_1-\tau_2-\tau_3)} d\tau_3 d\tau_2 d\tau_1 \delta(t-a_1)
\end{aligned}
$$

$$
\begin{aligned}
f_{3,2,1}(t) &= \int_0^{t} f_{2,1,1}(\tau_1) \lambda_{2,S_0-1} e^{-(\lambda_{2,S_0-1}+\mu_2)(t-\tau_1)} d\tau_1 \theta(a_2-t) \\
&+ \int_0^{t} f_{3,1,1}(\tau_1) \int_0^{t-\tau_1} \mu_3 e^{-(\lambda_{3,S_0-1}+\mu_3)\tau_2} \lambda_{2,S_0-1} e^{-(\lambda_{2,S_0-1}+\mu_2)(t-\tau_1-\tau_2)} d\tau_2 d\tau_1 \theta(a_2-t) \\
&+ \int_0^{a_1} f_{2,2,0}(\tau_1) e^{-(\lambda_{2,S_0-2}+\mu_2)(a_1-\tau_1)} d\tau_1 \delta(t-a_1) \\
&+ \int_0^{a_1} f_{3,2,0}(\tau_1) \int_0^{a_1-\tau_1} \mu_3 e^{-(\lambda_{3,S_0-2}+\mu_3)\tau_2} d\tau_2 d\tau_1 \delta(t-a_1)
\end{aligned}
$$

$$
\begin{aligned}
f_{4,2,1}(t) &= \int_0^{t} f_{3,1,1}(\tau_1) e^{-(\lambda_{3,S_0-1}+\mu_3)(t-\tau_1)} d\tau_1 \theta(a_2-t) \\
&+ \int_0^{a_1} f_{3,2,0}(\tau_1) e^{-(\lambda_{3,S_0-2}+\mu_3)(a_1-\tau_1)} d\tau_1 \delta(t-a_1)
\end{aligned}
$$

$$
\begin{aligned}
f_{2,3,0}(t) &= \int_0^{t} f_{2,2,0}(\tau_1) \int_0^{t-\tau_1} \mu_2 e^{-(\mu_2+\lambda_{2,S_0-2})\tau_2} \lambda_{1,S_0-2} e^{-(\lambda_{1,S_0-2}+\mu_1)(t-\tau_1-\tau_2)} d\tau_2 d\tau_1 \theta(a_1-t) \\
&+ \int_0^{t} f_{3,2,0}(\tau_1) \int_0^{t-\tau_1} \mu_3 e^{-(\mu_3+\lambda_{3,S_0-2})\tau_2} \int_0^{t-\tau_1-\tau_2} \mu_2 e^{-(\mu_2+\lambda_{2,S_0-2})\tau_3} \lambda_{1,S_0-2} e^{-(\lambda_{1,S_0-2}+\mu_1)(t-\tau_1-\tau_2-\tau_3)} d\tau_3 d\tau_2 d\tau_1 \theta(a_1-t)
\end{aligned}
$$

$$
\begin{aligned}
f_{3,3,0}(t) &= \int_0^{t} f_{2,2,0}(\tau_1) \lambda_{2,S_0-2} e^{-(\lambda_{2,S_0-2}+\mu_2)(t-\tau_1)} d\tau_1 \theta(a_1-t) \\
&+ \int_0^{t} f_{3,2,0}(\tau_1) \int_0^{t-\tau_1} \mu_3 e^{-(\mu_3+\lambda_{3,S_0-2})\tau_2} \lambda_{2,S_0-2} e^{-(\lambda_{2,S_0-2}+\mu_2)(t-\tau_1-\tau_2)} d\tau_2 d\tau_1 \theta(a_1-t)
\end{aligned}
$$

$$
f_{4,3,0}(t) = \int_0^{t} f_{3,2,0}(\tau_1) \lambda_{3,S_0-2} e^{-(\lambda_{3,S_0-2}+\mu_3)(t-\tau_1)} d\tau_1 \theta(a_1-t)
$$

Generalizing these equations, $f_{k,m,r}(t)$ for $m+r=1,2,3$, to the case of an arbitrary value of $m$ + $r$ is straightforward. The recursion relation satisfied by $f_{k,m,r}(t)$ is given by

$$
f_{k,m,r}(t)=\begin{pmatrix} \int_0^t f_{k-1,m-1,r}(\tau_1)\lambda_{k-1,S_0-(m-1)}e^{-(\lambda_{k-1,S_0-(m-1)}+\mu_{k-1})(t-\tau_1)}d\tau_1 \\ +\sum_{q=0}^{m+r-k}\int_0^t f_{k+q,m-1,r}\int_0^{t-\tau_1} g_{k,q,S_0,m}(\tau_2)\lambda_{k-1,S_0-(m-1)}e^{-(\lambda_{k-1,S_0-(m-1)}+\mu_{k-1})(t-\tau_1-\tau_2)}d\tau_2 d\tau_1 \end{pmatrix}\theta(a_{r+1}-t)
$$

$$
+\begin{pmatrix} \int_0^{a_r} f_{k-1,m,r-1}(\tau_1)e^{-(\lambda_{k-1,S_0-m}+\mu_{k-1})(a_r-\tau_1)}d\tau_1 \\ +\sum_{q=0}^{m+r-k}\int_0^{a_r} f_{k+q,m,r-1}(\tau_1)\int_0^{a_r-\tau_1} g_{k,q,S_0,m+1}(\tau_2)e^{-(\lambda_{k-1,S_0-m}+\mu_{k-1})(a_r-\tau_1-\tau_2)}d\tau_2 d\tau_1 \end{pmatrix}\delta(t-a_r).
$$

(A-3)

Based on Eq. (A-3), we can derive $P_n(t)$. It is useful to distinguish between the three $P_n(t)$ scenarios. First, we derive $P_n(t)$ when the infection count *n* ranges from 1 to $S_0+v$. For *n* = 1, we classify $P_1(t)$ into two groups based on infection, removal, or arrival occurrences. As only the index case exists at time zero, the number of infected individuals is one at time *t* if infection, removal, or arrival does not occur during *t*, which is written as $e^{-(\lambda_{1,S_0}+\mu_1)t}\theta(a_1-t)$. Next, we consider $P_1(t)$ in the presence of at least one infection, removal, or influx and assume that *k* infected individuals exist when the *m*th infection or the *r*th influx occurs at time $\tau_1$, and *k*-1 infected individuals die or recover during $\tau_2$. Then we can have one infected individual at time *t* unless infection and removal occur during $t-\tau_1-\tau_2$ and the (*r*+1)th arrival occurs during *t*. Mathematically, this is expressed as

$$
\sum_{m=0}^{S_0}\sum_{\substack{r=0 \\ m+r\geq 1}}^{v}\int_0^t f_{1,m,r}(\tau_1)e^{-(\lambda_{1,S_0-m}+\mu_1)(t-\tau_1)}d\tau_1\theta(a_{r+1}-t)
$$

$$
+\sum_{m=0}^{S_0}\sum_{\substack{r=0 \\ m+r\geq 1}}^{v}\sum_{k=2}^{m+r+1}\int_0^t f_{k,m,r}(\tau_1)\int_0^{t-\tau_1} h_{k,k-2,S_0}(\tau_2)e^{-(\lambda_{1,S_0-m}+\mu_1)(t-\tau_1-\tau_2)}d\tau_2 d\tau_1\theta(a_{r+1}-t).
$$

Therefore, we obtain

$$P_1(t) = e^{-(\lambda_{1,S_0}+\mu_1)t}\theta(a_1 - t) + \sum_{m=0}^{S_0}\sum_{\substack{r=0 \\ m+r\geq 1}}^{v}\int_0^t f_{1,m,r}(\tau_1)e^{-(\lambda_{1,S_0-m}+\mu_1)(t-\tau_1)}d\tau_1\theta(a_{r+1}-t)$$
$$+\sum_{m=0}^{S_0}\sum_{\substack{r=0 \\ m+r\geq 1}}^{v}\sum_{k=2}^{m+r+1}\int_0^t f_{k,m,r}(\tau_1)\int_0^{t-\tau_1} h_{k,k-2,S_0}(\tau_2)e^{-(\lambda_{1,S_0-m}+\mu_1)(t-\tau_1-\tau_2)}d\tau_2 d\tau_1\theta(a_{r+1}-t) \quad . \qquad \text{(A-4)}$$

For $n = 2$, a method similar to that used to derive Eq. (A-4) can be applied to the corresponding expression, $P_2(t)$. If *k* infected individuals are present when the *m*th infection or *r*th arrival occurs at time $\tau_1$, then $k-2$ infected individuals must be removed during $\tau_2$. When $k > 2$, there are two infected individuals at time *t* unless additional infection and removal events occur during $t-\tau_1-\tau_2$ and the (*r*+1)th arrival takes place before time *t*. This situation is described by the corresponding expression

$$P_2(t) = \sum_{m=0}^{S_0}\sum_{\substack{r=0 \\ m+r\geq 1}}^{v}\sum_{k=2}^{m+r+1}\int_0^t f_{k,m,r}(\tau_1)\left(\delta_{2,k}e^{-(\lambda_{2,S_0-m}+\mu_2)(t-\tau_1)} + (1-\delta_{2,k})\int_0^{t-\tau_1} h_{k,k-3,S_0}(\tau_2)e^{-(\lambda_{2,S_0-m}+\mu_2)(t-\tau_1-\tau_2)}d\tau_2\right)d\tau_1\theta(a_{r+1}-t)$$

. The derivation for general *n* follows in a similar manner:

$$P_n(t) = \sum_{m=0}^{S_0}\sum_{\substack{r=0 \\ m+r\geq n-1}}^{v}\sum_{k=n}^{m+r+1}\int_0^t f_{k,m,r}(\tau_1)\begin{pmatrix}\delta_{n,k}e^{-(\lambda_{n,S_0-m}+\mu_n)(t-\tau_1)} \\ +(1-\delta_{n,k})\int_0^{t-\tau_1} h_{k,k-(n+1),S_0}(\tau_2)e^{-(\lambda_{n,S_0-m}+\mu_n)(t-\tau_1-\tau_2)}d\tau_2\end{pmatrix}d\tau_1\theta(a_{r+1}-t) \qquad \text{(A-5)}$$

Eq. (A-5) can be simplified using Eq. (A-3), where where the first and last terms on the right-hand side of Eq. (A-3) are denoted accordingly

$$x_{k,m,r}(t) = \begin{pmatrix}\int_0^t f_{k-1,m-1,r}(\tau_1)\lambda_{k-1,S_0-(m-1)}e^{-(\lambda_{k-1,S_0-(m-1)}+\mu_{k-1})(t-\tau_1)}d\tau_1 \\ +\sum_{q=0}^{m+r-k}\int_0^t f_{k+q,m-1,r}\int_0^{t-\tau_1} g_{k,q,S_0,m}(\tau_2)\lambda_{k-1,S_0-(m-1)}e^{-(\lambda_{k-1,S_0-(m-1)}+\mu_{k-1})(t-\tau_1-\tau_2)}d\tau_2 d\tau_1\end{pmatrix}\theta(a_{r+1}-t)$$

and

$$y_{k,m,r}(t) = \begin{pmatrix}\int_0^{a_r} f_{k-1,m,r-1}(\tau_1)e^{-(\lambda_{k-1,S_0-m}+\mu_{k-1})(a_r-\tau_1)}d\tau_1 \\ +\sum_{q=0}^{m+r-k}\int_0^{a_r} f_{k+q,m,r-1}(\tau_1)\int_0^{a_r-\tau_1} g_{k,q,S_0,m+1}(\tau_2)e^{-(\lambda_{k-1,S_0-m}+\mu_{k-1})(a_r-\tau_1-\tau_2)}d\tau_2 d\tau_1\end{pmatrix}\delta(t-a_r) \, .$$

Compared with Eq. (A-3), Eq. (A-5) can be written as follows:

$$P_n(t)=\frac{1}{2}\sum_{m=0}^{S_0}\sum_{\substack{r=0\\m+r\geq n}}^{v}\left(\frac{x_{n+1,m,r}(t)}{\lambda_{n,S_0-(m-1)}}+y_{n+1,m,r}(t)\frac{\theta(a_r-t)}{\delta(t-a_r)}\right)\quad\text{for } n=2,3,\cdots,S_0+v \qquad \text{(A-6).}$$

Moreover, this work addresses $P_0(t)$, the probability of having no infected individuals at time *t*, from two perspectives. One perspective is that the index case is removed without new infections or arrivals. In this case, the probability that no infected individual exists at time *t* is $\int_0^t \mu_1 e^{-(\mu_1+\lambda_{1,S_0})\tau}d\tau\theta(a_1-t)$. The other perspective is that all infected individuals are removed when at least one infection or arrival occurs. If *k* infected individuals exist when *m* infections and *r* arrivals occur at time $\tau_1$, then the *k* infected cases must be removed during the interval $\tau-\tau_1$ for no infected individuals to exist at time $\tau$. The contribution of this case to $P_0(t)$ is given by $\int_0^t\int_0^\tau f_{k,m,r}(\tau_1)v_{k,m,S_0}(\tau-\tau_1)d\tau_1 d\tau\,\theta(a_{r+1}-t)$, where $v_{k,m,S_0}(t)$ represents the inverse Laplace transform of $\hat{v}_{k,m,S_0}(s)=\int_0^\infty v_{k,m,S_0}(t)e^{-st}dt=\prod_{l=1}^{k}\frac{\mu_l}{s+\mu_l+\lambda_{l,S_0-m}}$. Combining these two perspectives yields

$$\begin{aligned}P_0(t)&=\int_0^t\left(\mu_1 e^{-(\mu_1+\lambda_{1,S_0})\tau}\theta(a_1-t)+\sum_{m=0}^{S_0}\sum_{\substack{r=0\\m+r\geq 1}}^{v}\sum_{k=1}^{m+r+1}\left(\int_0^\tau f_{k,m,r}(\tau_1)v_{k,m,S_0}(\tau-\tau_1)d\tau_1\,\theta(a_{r+1}-t)\right)\right)d\tau\\&=\int_0^t\mu_1 P_1(\tau)d\tau\end{aligned}\quad,\quad\text{(A-7)}$$

where the last equality is obtained by the comparing of Eqs. (A-4) and (A-5).

The final point addresses the case on which the number of infections takes the maximum value $S_0+v+1$. For the number of infected individuals to reach the maximum value, all susceptible individuals must become infected and as many infected individuals as possible must arrive. The number of infected individuals is $S_0+v+1$ at time $\tau$, and recovery or death must not occur during the time interval $t-\tau$. We thus have

$$P_{S_0+v+1}(t)=\int_0^t f_{S_0+v+1,S_0+1,v}(\tau)e^{-\mu_{S_0+v+1}(t-\tau)}d\tau\quad. \qquad \text{(A-8)}$$

Combining Eqs. (A-6) through (A-8) yields Eq. (8). We verify the normalization condition $\sum_{n=0}^{S_0+v+1} P_n(t)=1$ by rearranging the sum of $P_n(t)$ from $n = 0$ to $S_0+v+1$:

$$\sum_{n=0}^{S_0+v+1} P_n(t)=1-\sum_{m=0}^{S_0}\sum_{\substack{r=0\\ m+r=1}}^{v}\sum_{k=1}^{m+r+1}\int_0^t f_{k,m,r}(\tau)d\tau+\left(\sum_{m=0}^{S_0}\sum_{\substack{r=0\\ m+r=1}}^{v}\sum_{k=1}^{2}\int_0^t f_{k,m,r}(\tau)d\tau-\sum_{m=0}^{S_0}\sum_{\substack{r=0\\ m+r=2}}^{v}\sum_{k=1}^{3}\int_0^t f_{k,m,r}(\tau)d\tau\right)$$
$$+\left(\sum_{m=0}^{S_0}\sum_{\substack{r=0\\ m+r=2}}^{v}\sum_{k=1}^{3}\int_0^t f_{k,m,r}(\tau)d\tau-\sum_{m=0}^{S_0}\sum_{\substack{r=0\\ m+r=3}}^{v}\sum_{k=1}^{4}\int_0^t f_{k,m,r}(\tau)d\tau\right)+\cdots$$
$$+\left(\sum_{m=0}^{S_0}\sum_{\substack{r=0\\ m+r=S_0+v-1}}^{v}\sum_{k=1}^{S_0+v}\int_0^t f_{k,m,r}(\tau)d\tau-\sum_{m=0}^{S_0}\sum_{\substack{r=0\\ m+r=S_0+v}}^{v}\sum_{k=1}^{S_0+v+1}\int_0^t f_{k,m,r}(\tau)d\tau\right)+\sum_{m=0}^{S_0}\sum_{\substack{r=0\\ m+r=S_0+v}}^{v}\sum_{k=1}^{S_0+v+1}\int_0^t f_{k,m,r}(\tau)d\tau=1$$

(A-9)

If infected individuals do not arrive, Eqs. (A-6) to (A-8) align with Eq. (1) since $v = 0$.

The Laplace transform of Eq. (8) allows one to obtain the differential-difference equation for $P_n(t)$, given in Eq. (9). Specifically, the Laplace transform of $P_1(t)$ becomes $s\hat{P}_1(s)-1=-\sum_{m=0}^{S_0}\sum_{r=0}^{v}\hat{f}_{2,m,r}(s)-\mu_1\hat{P}_1(s)+\mu_2\hat{P}_2(s)$ because of the initial condition $P_n(0)=\delta_{n1}$, where $\hat{P}_n(s)=\int_0^\infty P_n(t)e^{-st}dt$. Moreover, $\hat{P}_2(s)$ yields $s\hat{P}_2(s)=-\sum_{m=0}^{S_0}\sum_{\substack{r=0\\ m+r\ge 2}}^{v}\hat{f}_{3,m,r}(s)-\mu_2\hat{P}_2(s)+\sum_{m=0}^{S_0}\sum_{\substack{r=0\\ m+r\ge 1}}^{v}\hat{f}_{2,m,r}(s)+\mu_3\hat{P}_3(s)$. Continuing to $\hat{P}_3(s),\hat{P}_4(s),\cdots$, we obtain $s\hat{P}_n(s)=-\sum_{m=0}^{S_0}\sum_{\substack{r=0\\ m+r\ge n}}^{v}\hat{f}_{n+1,m,r}(s)-\mu_n\hat{P}_n(s)+\sum_{m=0}^{S_0}\sum_{\substack{r=0\\ m+r\ge n-1}}^{v}\hat{f}_{n,m,r}(s)+\mu_{n+1}\hat{P}_{n+1}(s)$ for $n=2,3,\cdots,S_0+v$. From Eqs. (A-7) and (A-8), we have $s\hat{P}_0(s)=\mu_1\hat{P}_1(s)$ and $s\hat{P}_{S_0+v+1}(s)=\hat{f}_{S_0+v+1,S_0+1,v}(s)\left(1-\frac{\mu_{S_0+v+1}}{s+\mu_{S_0+v+1}}\right)$. Taking the inverse Laplace transform of $\hat{P}_n(s)$ for $n=0,1,2,\cdots,S_0+v+1$ completes the proof of Eq. (9).